\documentclass[reprint, aps, pra]{revtex4-2}
\usepackage{amsfonts}
\usepackage{amsmath}
\usepackage{amssymb}
\usepackage{mathtools}
\usepackage{braket}
\usepackage{bm}
\usepackage{dsfont}

\usepackage{graphicx}
\usepackage{float}
\usepackage{epstopdf}

\usepackage{hyperref}
\usepackage{dcolumn}

\usepackage{color}
\usepackage[usenames, dvipsnames]{xcolor}

\hypersetup{
  colorlinks=true,
  linkcolor=RoyalBlue,
  linkbordercolor=RoyalBlue,
  citecolor=Green,
  citebordercolor=Green,
  urlcolor=Cerulean,
  urlbordercolor=Cerulean
}

\newcommand{\ddt}[1][]{%
  \frac{\mathrm{d} #1}{\mathrm{d}t}
}

\begin{document}

\preprint{APS/123-QED}

%==============================================================================%
%                                 FRONTMATTER                                  %
%==============================================================================%

\title{Multi-Mode Array Filtering of Resonance Fluorescence}

\author{Jacob Ngaha}
\email{j.ngaha@auckland.ac.nz}
\author{Scott Parkins}
\author{Howard J. Carmichael}
\affiliation{The Dodd-Walls Centre for Photonic and Quantum Technologies, Department of Physics \\ University of Auckland, Private Bag 92019, Auckland, New Zealand}

\date{\today}

%==============================================================================%
%                                  ABSTRACT                                    %
%==============================================================================%
\begin{abstract}
  We present a frequency-filtering method for measuring and calculating frequency-filtered photon-correlations. This method is a cavity-based system we call the \emph{multi-mode array filter}, which consists of an array of tunable single-mode cavities that are equally spaced in frequency. By introducing a mode-dependent phase modulation, we produce a near rectangular frequency response, allowing us to increase the filter bandwidth -- and thus the temporal response -- without sacrificing frequency isolation. We model the frequency filtering using a cascaded quantum open systems approach which completely neglects any back-action of the filter onto the source system. This allows us to derive a closed set of operator moment equations for source and filter system operators, thus providing an extremely efficient method to calculate frequency-filtered first- and second-order correlation functions. We demonstrate this filtering method by applying it to a resonantly driven two-level atom. We present examples of frequency-filtered power spectra to demonstrate the improved frequency isolation of the multi-mode array filter over the single-mode filter. We then present results for the single-mode and multi-mode-array filtered second-order auto- and cross-correlation functions. These are compared against expressions derived in the secular approximation. The improved frequency isolation of the multi-mode array filter allows us to investigate new regimes of frequency-filtered photon correlations, such as two-photon \emph{leapfrog} processes, and the effect of vanishing bandwidth on filtered auto-correlation functions.
\end{abstract}

%Use showkeys class option if keyword display

% Make title display
\maketitle

%==============================================================================%
%                                1 INTRODUCTION                                %
%==============================================================================%
\section{Introduction}

Resonance fluorescence of a driven two-level atom is one of the simplest examples of light interacting with matter, yet it has also played a hugely important role in the development of quantum optics. It was found that, upon strong coherent excitation, the fluorescence spectrum splits into three components, known as the ``Mollow triplet'' \cite{Mollow_1969_PR_PowerSpectrumLight, Mollow_1975_PRA_PureStateAnalysis}. Around the same time of Mollow's work, there were also investigations into the quantum nature of the emitted light, based on Glauber's photon correlation work \cite{Glauber_1963_PR_QuantumTheoryOptical, Glauber_1963_PRL_PhotonCorrelations}. It was shown that the emitted photons are \textit{antibunched} -- an entirely quantum effect \cite{Carmichael_1976_JoPBAaMP_ProposalMeasurementResonant, Carmichael_1976_JoPBAaMP_QuantumMechanicalMaster, Walls_1979_N_EvidenceQuantumNature, Kimble_1977_PRL_PhotonAntibunchingResonance}.

The majority of studies on atomic resonance fluorescence at this time were largely focused on the entire fluorescence spectrum of the atom. It was at the end of the 1970s when some of the first investigations were conducted on correlations between photons from different components of the Mollow triplet, as opposed to the whole spectrum \cite{Apanasevich_1977_PLA_LightInducedCorrelations, Apanasevich_1978_JoAS_MultitimeCorrelationScattering, Apanasevich_1979_JoPBAaMP_PhotonBunchingAntibunching, Arnoldus_1984_JoPBAaMP_PhotonCorrelationsLines, Cresser_1987_JoPBAaMP_IntensityCorrelationsFrequency}, with the first experimental results reported by Aspect, Roger, Dalibard and Cohen-Tannoudji \cite{Aspect_1980_PRL_TimeCorrelationsTwo}. To achieve the separation of the two side-peaks, the atomic fluorescence was split into two channels, with each channel passing through an interference filter tuned to the respective target frequency, and modelled as a single-mode optical cavity \cite{Eberly_1977_JotOSoA_TimeDependentPhysical, Knoell_1984_JoPBAaMP_TheoryTimeResolved, Knoell_1990_PRA_SpectralPropertiesLight}. Filtering such as this is essential to the measurement of frequency-resoled photon correlations as, typically, photon detectors are broad bandwidth, and essentially non-selective in frequency.

With this increased interest in measuring photon correlations of frequency components came interest in the effect that spectral filtering had on measurements \cite{Nienhuis_1993_PRA_SpectralCorrelationsResonance, Nienhuis_1993_EPL_SpectralCorrelationsFluorescence, Joosten_2000_JoOBQaSO_InfluenceSpectralFiltering, Cresser_1983_PR_TheorySpectrumQuantised, Schmidt_2021_QSaT_FrequencyResolvedPhoton}, as well as interest in other methods of calculating and measuring these correlations. In 2012, del Valle et al. proposed a novel method for measuring frequency-resolved photon correlations \cite{Valle_2012_PRL_TheoryFrequencyFiltered}, consisting of a system of interest weakly coupled to $N$ two-level ``detector'' atoms, where $N$ is the order of the correlation function of interest. This method has been used extensively in theoretical studies due to its relatively simple implementation \cite{Valle_2013_NJoP_DistillingOneTwo, GonzalezTudela_2013_NJoP_TwoPhotonSpectra, Munoz_2014_PRA_ViolationClassicalInequalities, GonzalezTudela_2015_PRA_OptimizationPhotonCorrelations, Munoz_2015_NJoP_EnhancedTwoPhoton, Darsheshdar_2021_PRA_PhotonPhotonCorrelations, Laussy_2017_NM_NewWayCorrelate}. There have been other methods discussed over the years, some with more theoretical leaning, such as: signal processing methods \cite{Shatokhin_2016_PRA_CorrelationFunctionsResonance}; perturbation approaches \cite{Holdaway_2018_PRA_PerturbationApproachComputing}; frequency resolved Monte-Carlo, or quantum trajectory methods \cite{Holland_1998_PRL_UnravelingQuantumDissipation, Carreno_2018_SR_FrequencyResolvedMonte, Bel_2009_PRL_TheoryWavelengthResolved}. Some other methods, such as the eigenvalue decomposition of Liouvillian superoperators developed by Kamide et al \cite{Kamide_2015_PRA_EigenvalueDecompositionMethod, Clauser_1971_PRB_RelaxationEffectsSpectra}, allow for the modelling of not only Lorentzian type filters, but for filters with any frequency response.

While there is no single method of frequency filtering that is ideal for every application, there are also some issues with the previously mentioned methods. The most common, and perhaps most intuitive, method of frequency filtering is the tunable interferometer. The main downside to this method, however, is the Lorentzian frequency response of the filter -- a response that is also shared with the detector atom approach -- due to the far reaching tails of the Lorentzian distribution which, in fact, never decay entirely. For multichromatic sources, such as the Mollow triplet, the tails of the frequency response can readily overlap with non-target frequency components. As the bandwidth of a frequency filter is inversely proportional to its temporal response, and faster temporal responses are required for accurate correlation measurements, there will always be a trade-off between frequency isolation and temporal resolution with a Lorentzian filter. A sought-after aim, therefore, is to develop a realistic frequency filter model with a much sharper frequency response than a standard Lorentzian, such as a rectangular filter.

We also desire that the filter has no effect on the evolution of the source system. This is usually achieved by assuming a vanishingly small coupling of the source system to the filter, such that any back-action can be neglected. Holdaway et al. \cite{Holdaway_2018_PRA_PerturbationApproachComputing} take this approximation a step further with an algebraic expansion of the source-filter system with respect to the coupling parameter. We can, however, ensure there is no back-action at all by \textit{cascading} the output of the source system into the filter, using cascaded open quantum systems theory \cite{Gardiner_1993_PRL_DrivingQuantumSystem, Carmichael_1993_PRL_QuantumTrajectoryTheory}. This can be achieved experimentally through the use of forward propagating waveguides, ring cavities, optical isolators, or beam splitters.

In this paper, we will introduce our solution to these problems: the \textit{multi-mode array filter}, which consists of an array of tunable, single-mode cavities. The output of a source system is cascaded equally into each mode, where a mode-dependent phase modulation is applied. The outputs of all modes are then combined, allowing an interference which results in an approximately rectangular frequency response. To ensure a rectangular frequency response, we consider a large number of modes in the array, and thus calculations can be very computationally expensive. We therefore employ a further, uncommon approach to calculating these frequency-filtered second-order correlation functions, by deriving rate equations for the expectation values of the detection operators. With no back-action of the array filter on the source system, these operator -- or moment -- equations, form an effective coupling hierarchy where higher order moments are only dependent on lower orders. We can then solve for exact solutions of the moment equations extremely efficiently.

We begin in Sect.~\ref{sect:2_filtering} by introducing the foundations of our filtering model, including a brief discussion of single-mode cavities and rectangular filters. We then present analytic and numerical results for the temporal and frequency response of the multi-mode array filter for different parameters. In Sect.~\ref{sect:3_atom} we review the dynamics and key phenomena of a resonantly driven two-level atom, which will serve as the source system with which to test the effectiveness of the filter. We then introduce the full atom-filter coupled system in Sect.~\ref{sect:4_moments_g1} and the moment equations relevant to solving for the two-time correlation functions. In Sect.~\ref{sect:6_photon_correlations} we present analytic results for frequency-filtered auto- and cross-correlations, comparing the multi-mode array-filtered results with those of a standard Lorentzian filter and approximate ideal solutions derived in the dressed-state basis. Finally, we conclude in Sect.~\ref{sect:6_conclusion} with a summary of key results and a brief discussion on possible experimental implementations.

%==============================================================================%
%                          2 MULTI-MODE ARRAY FILTER                           %
%==============================================================================%
\section{Frequency Filtering Through Optical Cavities}
\label{sect:2_filtering}

%------------------------------------------------------------------------------%
%-------------------------------------------%
%     2.1 Single-mode Lorentzian filter     %
%-------------------------------------------%
\subsection{Single-mode Lorentzian filter}
%------------------------------------------------------------------------------%

We begin by considering a standard frequency filtering device, modelled as a single-mode cavity. Applying a coherent driving, with driving strength $\mathcal{E}_{d}$, the Hamiltonian modelling this system is
\begin{equation}\label{eq:2_hamiltonian_single_mode_filter}
  H_{F} = \hbar \omega_{c} a^{\dagger} a + i \hbar \left( \mathcal{E}_{d} e^{-i \omega t} a^{\dagger} - \mathcal{E}_{d}^{*} e^{i \omega t} a \right) ,
\end{equation}
where $\omega_{c}$ is the central resonance frequency of the cavity mode, $\omega$ is the frequency of the driving field, and $a^{\dagger}~\left( a \right)$ is the photon creation~(annihilation) operator for the single-cavity mode. We account for photon loss, at rate $2 \kappa$, with the quantum optical master equation for the cavity mode density operator $\rho$,
\begin{equation}
  \ddt[\rho] = \frac{ 1 }{ i \hbar } [ H_{F}, \rho ] + \kappa \Lambda(a) \rho ,
\end{equation}
with Lindblad superoperator,
\begin{equation}
  \Lambda(X)\bullet = \left( 2 X \bullet X^{\dagger} - X^{\dagger} X \bullet - \bullet X^{\dagger} X \right) .
\end{equation}
We assume the cavity to initially be in the vacuum state, and derive the equation of motion for the coherent state amplitude of the cavity mode as
\begin{equation}\label{eq:2_single_mode_alpha_rate}
  \ddt \alpha = - \left( \kappa + i \omega_{c} \right) \alpha + \mathcal{E}_{d} e^{i \omega t} ,
\end{equation}
which, in the long time limit $t \rightarrow \infty$, has solution
\begin{equation}\label{eq:2_single_mode_amplitude}
  \alpha(\omega, t) = \frac{ \mathcal{E}_{d} e^{-i \omega t }}{\kappa - i \left( \omega - \omega_{c} \right) }.
\end{equation}
The frequency response of the single-mode cavity is, as expected, a simple Lorentzian, with halfwidth $\kappa$,
\begin{equation}\label{eq:2_single_mode_frequency_response}
  |\alpha(\omega)|^{2} = \frac{|\mathcal{E}_{d}|^{2}}{\kappa^{2} + \left( \omega - \omega_{c} \right)^{2}} .
\end{equation}
We therefore arrive at the main issue with single-mode cavities: the Lorentzian distribution. The filter bandwidth, $2 \kappa$, is inversely proportional to the mean lifetime of a photon inside the cavity. A faster temporal response, which is ideal for measuring accurate photon correlations, therefore requires a larger bandwidth, yet the tails of the Lorentzian distribution extend far from the central frequency. If the source system of interest emits photons over a range of frequencies, as we consider in this paper, the frequency response of a single-mode filter may overlap with non-target frequencies. Decreasing the bandwidth to reject these non-target frequencies then increases the mean photon lifetime, resulting in less accurate frequency-filtered photon correlations.

%------------------------------------------------------------------------------%
%------------------------------------%
%     2.2 The rectangular filter     %
%------------------------------------%
\subsection{The rectangular filter}
%------------------------------------------------------------------------------%

\begin{figure}
  \centering
  \includegraphics[width=\linewidth]{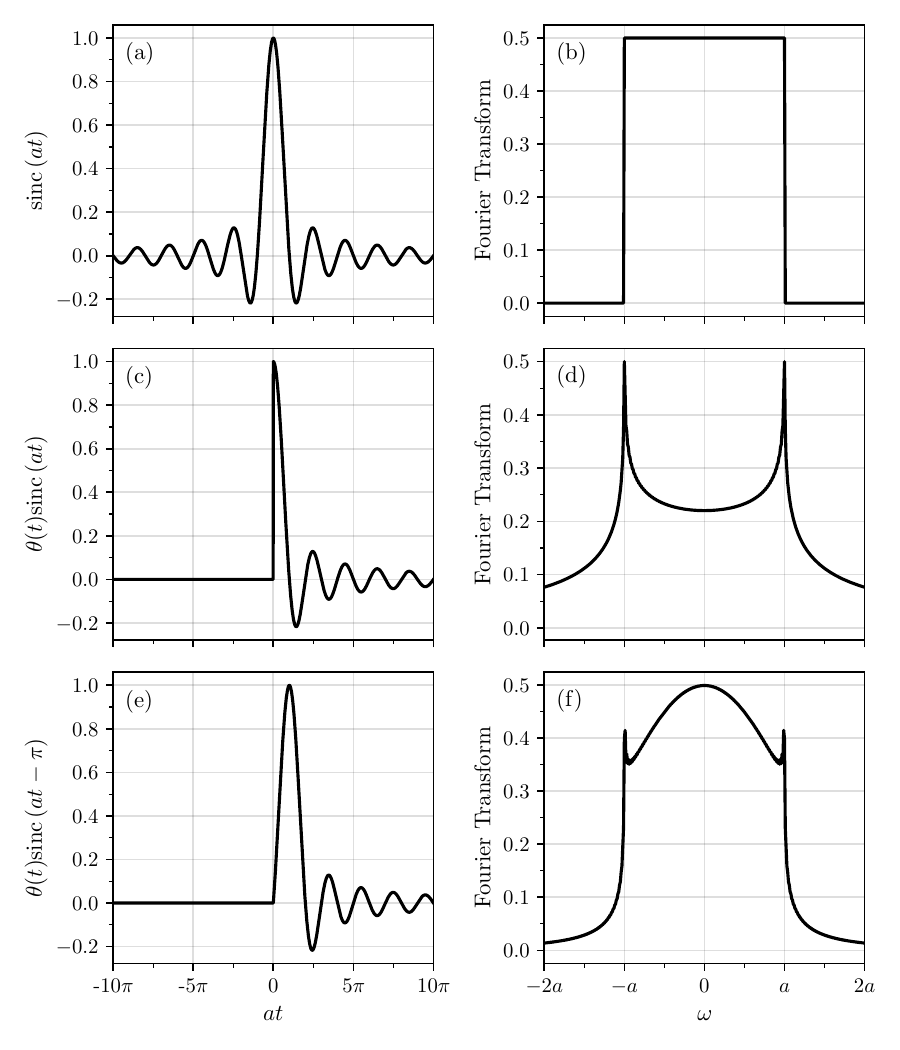}
  \caption{Time-series and Fourier transform of a complete sinc function (a, b), positive-side sinc function (c, d), and a positive sinc function with a $\pi$ phase delay (e, f).}
  \label{fig:2_sinc_stuff}
\end{figure}

Our ultimate goal is to develop a cavity-based frequency filter that allows us to pick specific frequency components of a general source system's spectrum and measure photon correlations. We take inspiration from an ideal bandpass model, i.e., the sinc filter -- also known as the rectangular filter  -- which results from the Fourier transform of the sinc function in the time domain:
\begin{align}
  \mathrm{Rect}(\omega) &= \frac{ a }{ 2 \pi } \int_{-\infty}^{\infty} \frac{ \mathrm{sin} \left( a t \right) }{ a t } e^{i \omega t } \mathrm{d}t \nonumber \\
  &=
  \begin{cases}
    0 &\quad a < | \omega | \\
    \frac{1}{4} &\quad | \omega | = a \\
    \frac{1}{2} &\quad | \omega | < a \\
  \end{cases} ,
\end{align}
as shown in Figs.~\ref{fig:2_sinc_stuff}(a) and (b). The infinitely sharp frequency response of the sinc, or rectangular, filter, would allow us to maximally increase the bandwidth, and thus the temporal response, of the filter response, much more than a conventional Lorentzian filter, while still maintaining good frequency isolation. Unfortunately, perfect sinc filters are physically impossible to realise due to their \textit{non-causal} nature, that is, the temporal response depends on future inputs. By neglecting the negative temporal response of the sinc function with $\theta(t) \mathrm{sinc}( a t )$, where
\begin{equation}
  \theta(t) =
  \begin{cases}
    0 &\quad t \leq 0 \\
    1 &\quad 0 < t
  \end{cases}
\end{equation}
is the Heaviside step function, we can realise a causal sinc filter. We see in Fig.~\ref{fig:2_sinc_stuff}(d), however, that the frequency response, as calculated from the Fourier transform, changes drastically. The infinitely sharp sides of the rectangular response have been replaced with large slow-decaying wings, together with a large dip in the centre of the response.

Fortunately we are able to recover some of the rectangular nature by introducing a slight phase modulation, or delay, to the temporal response, $\theta(t) \mathrm{sinc}( a t - \pi )$ (Fig.~\ref{fig:2_sinc_stuff}(e)). With this slight delay, the Fourier transform, Fig.~\ref{fig:2_sinc_stuff}(f), shows a much sharper cut-off than the non-delayed response in Fig.~\ref{fig:2_sinc_stuff}(d). By increasing the delay even further, we can ``recover'' much more of the sinc function, resulting in an even more rectangular frequency response. Unfortunately, a greater modulation also results in a greater temporal delay, which is undesirable for measuring accurate photon correlations. We will see in the following section that we are able to recreate this behaviour with a cavity based model.

%------------------------------------------------------------------------------%
%-------------------------------------%
%     2.3 Multi-Mode Array Filter     %
%-------------------------------------%
\subsection{The multi-mode array filter}
\label{sect:2_multi_mode_array_filter}
%------------------------------------------------------------------------------%

\begin{figure}
  \centering
  \includegraphics[width=\linewidth]{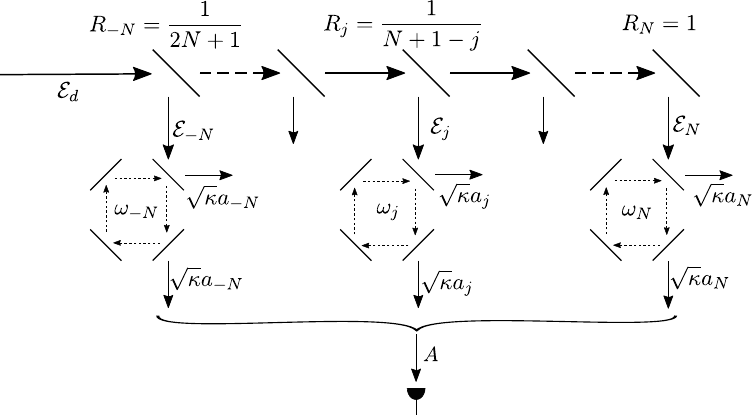}
  \caption{Schematic of a coherently driven multi-mode array filter mode. The input field is evenly split into each individual two-sided cavity mode, where a mode-dependent phase modulation is applied. To achieve this equal coupling of the input light, the input field is passed through an array of beam splitters with increasing intensity reflectivity coefficient. The other (vertical) input field to each beam splitter in the top row is vacuum. Similarly, the input field (horizontal) to the lower output mirror in each filter cavity is also vacuum. Since we consider photon counting measurements, these vacuum fields make no contribution to the detection process.}
  \label{fig:2_multi_mode_schematic}
\end{figure}

It is no simple feat to develop an optical cavity-based frequency filter that does not have a Lorentzian frequency response. We therefore propose a model where an input source field is coupled into an \textit{array of single-mode cavities}, each with a small bandwidth. At first glance, the small bandwidth of each individual cavity mode might appear to result in an overall poor temporal response of the filter as a whole, however we will show that this is not the case.

We consider a configuration of $2N + 1$ beam splitters and $2N + 1$ single-mode cavities, as illustrated in Fig.~\ref{fig:2_multi_mode_schematic}. A portion of the input field is reflected at each beam splitter towards the in-coupling mirror of a single-mode cavity. The resonance frequency of each cavity mode is detuned by a certain amount from some central frequency $\omega_{c}$. To achieve equal coupling of a source field into each cavity mode, the beam splitters have increasing reflectivity, $R_{j} = \left( N + 1 - j \right)^{-1}$. We therefore extend the single-mode Hamiltonian, Eq.~(\ref{eq:2_hamiltonian_single_mode_filter}), to the multi-mode array filter, with
\begin{multline}\label{eq:2_multi_mode_array_hamiltonian}
  H = \hbar \sum_{j=-N}^{N} \omega_{j} a_{j}^{\dagger} a_{j} \\
  + i \hbar \sum_{j=-N}^{N} \left( \mathcal{E}_{j} e^{-i \omega t} a_{j}^{\dagger} - \mathcal{E}_{j}^{*} e^{i \omega t} a_{j} \right),
\end{multline}
where $N$ is the number of cavity modes with frequencies on either side of the central mode frequency $\omega_{c}$, $a_{j}^{\dagger}~\left( a_{j} \right)$ is the photon creation~(annihilation) operator for the $j^{\mathrm{th}}$ filter mode, $\omega_{j} = \omega_{c} + j \delta\omega$ is the resonance frequency of the $j^{\mathrm{th}}$ filter mode with mode frequency spacing $\delta\omega$, and
\begin{equation}
  \mathcal{E}_{j} = \frac{ \mathcal{E}_{d} }{ \sqrt{ 2 N + 1} } e^{i m j \pi / N}
\end{equation}
is the mode-dependent driving amplitude. Note that the input field coupled into each filter mode has a mode-dependent phase modulation applied to it, where $m$ sets the size of the introduced modulation. Each individual mode has a total photon loss rate of $2 \kappa$, such that the master equation for the total system is:
\begin{equation}\label{eq:2_multi_mode_array_master_equation}
  \ddt[\rho] = \frac{ 1 }{ i \hbar } [ H, \rho ] + \kappa \sum_{j=-N}^{N} \Lambda \left( a_{j} \right) \rho .
\end{equation}
We then assume that the fields transmitted through the out-coupling mirrors of the cavities are incident on a single photon detector. Note that the input fields to these mirrors are in the vacuum state. As we are measuring the output via photon detectors, these vacuum input fields make no contribution to the detection process. The total electric field at the detector is then given by
\begin{equation}
  E^{(+)}_{\mathrm{source}}(z, t) = \sum_{j=-N}^{N} \sqrt{ \frac{ \hbar \omega_{j} }{ 2 \epsilon_{0} A c } } \frac{ \sqrt{ 2 \kappa } a_{j}(t - z/c) }{ \sqrt{ 2N + 1 } } ,
\end{equation}
where $\omega_{j}$ is the frequency of the $j^{\mathrm{th}}$ cavity mode, $\epsilon_{0}$ is the electric permittivity of free space, and $A$ is the quantisation area of the field. With frequencies in the optical range ($\sim 10^{15}$ Hz), we assume that the effective halfwidth of the multi-mode array filter, $N \delta\omega$, is much smaller than the central mode frequency $\omega_{0}$. The total field may then be approximated as a sum over the annihilation operators:
\begin{align}
  E^{(+)}_{\mathrm{source}}(z, t) &= \sqrt{ \frac{ \hbar \omega_{0} }{ 2 \epsilon_{0} A c } } \sqrt{ 2 \kappa } \sum_{j=-N}^{N} \frac{ a_{j}(t - z/c) }{ \sqrt{ 2N + 1} } \nonumber \\
  &\propto \bar{A}(t - z/c),
\end{align}
where
\begin{equation}
  \bar{A} = \frac{ 1 }{ \sqrt{ 2 N + 1 } } \sum_{j = -N}^{N} a_{j}
\end{equation}
is the \textit{normalised collective mode annihilation operator}. For the single-mode case, $N = 0$, $\kappa$ is also the halfwidth of the filter response. We therefore define the ``effective halfwidth'' of the multi-mode array filter as
\begin{equation}\label{eq:2_multi_mode_array_bandwidths}
  K =
  \begin{cases}
    \kappa &\quad N = 0 \\
    N \delta\omega &\quad N > 0
  \end{cases} .
\end{equation}

%------------------------------------------------------------------------------%
%---------------------------------%
%     2.3.1 Temporal response     %
%---------------------------------%
\subsubsection{Temporal response}
%------------------------------------------------------------------------------%

\begin{figure}
  \centering
  \includegraphics[width=\linewidth]{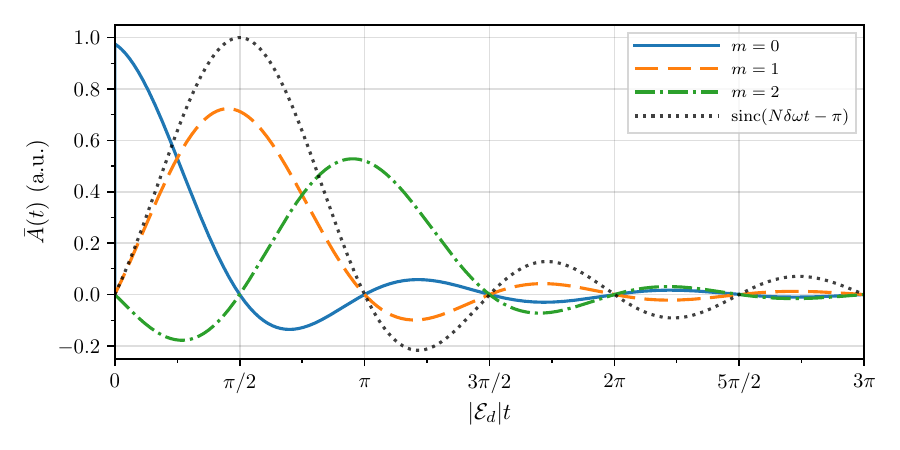}
  \caption{Normalised field amplitude response to an impulse driving, Eq.~(\ref{eq:2_multi_mode_array_temporal_response}), with $m=0$ (blue, solid), $m=1$ (orange, dashed), and $m=2$ (green, dash-dot). Also shown is the sinc term of Eq.~(\ref{eq:2_multi_mode_array_temporal_response}) (grey, dotted), showing the effect that the cavity decay, $\kappa$, has on the temporal response. Other parameters are $N = 20$ and $\left( \delta\omega, \kappa, \omega_{c} \right) = \left( 0.1, 0.2, 0.0 \right) |\mathcal{E}_{d}|.$}
  \label{fig:2_impulse_driving_cavity_amplitude}
\end{figure}

Before assessing the frequency response of the multi-mode array filter, we will first investigate the temporal response of the filter to an impulse driving. From Eqs.~(\ref{eq:2_multi_mode_array_hamiltonian}) and (\ref{eq:2_multi_mode_array_master_equation}) we can write an equation of motion for the coherent field amplitude of the $j^{\mathrm{th}}$ mode, however we replace the continuous driving term in Eq.~(\ref{eq:2_single_mode_alpha_rate}) with the Dirac delta-function, such that the equation of motion is
\begin{equation}
  \ddt \alpha_{j} = -\left( \kappa + i \omega_{j} \right) \alpha_{j} + \mathcal{E}_{j} \delta(t) ,
\end{equation}
with solution
\begin{equation}\label{eq:2_multi_mode_alpha_impulse}
  \alpha_{j}(t) = \mathcal{E}_{j} e^{-\left( \kappa + i \omega_{j} \right) t } \theta(t) ,
\end{equation}
where $\theta(t)$ is the Heaviside step-function. To find the total response of the multi-mode array filter, we sum all of the mode amplitudes and consider the collective field amplitude,
\begin{equation}\label{eq:2_multi_mode_temporal_amplitude}
  \bar{A}(t) = \sum_{j = -N}^{N} \frac{ \alpha_{j}(t) }{ \sqrt{ 2 N + 1 } } .
\end{equation}
Assuming a large number of closely spaced modes, such that $N \gg 1$ and $\delta\omega \ll \mathcal{E}_{d}$, we convert the summation in Eq.~(\ref{eq:2_multi_mode_temporal_amplitude}) into an integral, and thus find the collective field amplitude to be:
\begin{align}\label{eq:2_multi_mode_array_temporal_response}
  \bar{A}(t) &\simeq \frac{ 1 }{ \sqrt{ 2 N + 1 } } \int_{-\infty}^{\infty} \alpha(j, t) \mathrm{d} j \nonumber \\
  &= \frac{ 2 N \mathcal{E}_{d} }{ 2 N + 1 } ~ \theta(t) ~ e^{-i \left( \kappa + i \omega_{c} \right) t} \nonumber \\
  &\quad\quad\quad\quad \times \mathrm{sinc} \left( N \delta\omega t - m \pi \right) .
\end{align}
We see, then, that this proposed model behaves as we expect from a sinc-response filter, i.e., the temporal response is a positive-sided sinc function, with halfwidth $N \delta\omega$ and a temporal delay of $m \pi$. There is, however, an extra term that arises due to individual cavity modes which make up the multi-mode array filter: the exponential decay and oscillatory term $e^{-\left( \kappa + i \omega_{c}\right) t}$. The oscillatory term, $e^{-i \omega_{c}t}$, simply centres the response in frequency space at $\omega_{c}$. The cavity decay term, however, can have a much more significant effect on the filter's response. We demonstrate this in Fig.~\ref{fig:2_impulse_driving_cavity_amplitude}, where the temporal response for the phase-shifted multi-mode array (orange, dashed), with $m = 0, 1$, or 2, is plotted against the sinc term in Eq.~(\ref{eq:2_multi_mode_array_temporal_response}). The cavity decay causes the collective temporal response to deviate from the pure sinc term; an effect which is much more noticeable for larger $\kappa$. We can mitigate this effect, and thus achieve a more ideal sinc response, by ensuring the individual cavity decay rate is much smaller than the effective halfwidth of the array filter, i.e., $\kappa \ll N \delta\omega$.

%------------------------------------------------------------------------------%
%----------------------------------%
%     2.3.2 Frequency response     %
%----------------------------------%
\subsubsection{Frequency response}
%------------------------------------------------------------------------------%

\begin{figure}
  \centering
  \includegraphics[width=\linewidth]{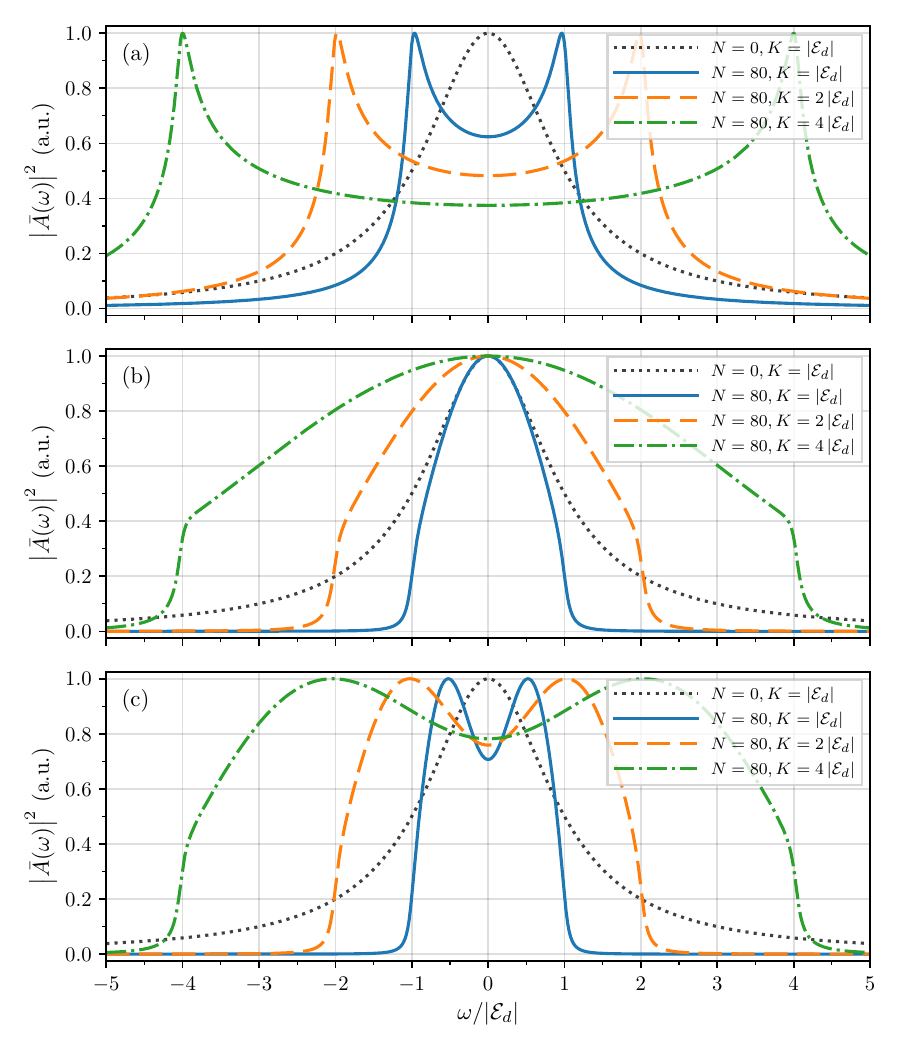}
  \caption{Normalised frequency response for a single-mode cavity with $\kappa = |\mathcal{E}_{d}|$ (grey, dotted) and a multi-mode array filter with (a) no phase modulation $(m=0)$, from Eq.~(\ref{eq:2_multi_mode_response_m0}), (b) $\pi$-phase modulation $(m=1)$, and (c) $2 \pi$-phase modulation $(m=2)$, with $K = |\mathcal{E}_{d}|$ (blue, solid), $K = 2 |\mathcal{E}_{d}|$ (orange, dashed), and $K = 4 |\mathcal{E}_{d}|$ (green, dot-dash). Other parameters are $N = 80, \omega_{c} = 0$, and, for the multi-mode array filter, $\kappa = 0.07 |\mathcal{E}_{d}|$.}
  \label{fig:2_frequency_response}
\end{figure}

The sinc-like temporal response of the multi-mode array filter is promising, yet it does not tell the entire story. To investigate the frequency response of this filter model, we must return to the case of continuous driving. With the same approach that led to Eq.~(\ref{eq:2_single_mode_frequency_response}), we write the rate equation for the mean cavity field amplitude of the $j^{\mathrm{th}}$ mode as
\begin{equation}
  \ddt \alpha_{j} = -\left( \kappa + i \omega_{j} \right) \alpha_{j} + \mathcal{E}_{j} e^{-i \omega t},
\end{equation}
which has solution in the long-time limit similar to Eq.~(\ref{eq:2_single_mode_amplitude}),
\begin{equation}\label{eq:2_multi_mode_alpha_continuous}
  \alpha_{j}(\omega, t) = \frac{ \mathcal{E}_{j} e^{-i \omega t} }{ \kappa - i \left( \omega - \omega_{j} \right) }.
\end{equation}
As with the temporal response, the collective frequency response of the multi-mode array filter is the sum of all field amplitudes:
\begin{equation}
  \bar{A}(\omega) = \frac{ 1 }{ \sqrt{ 2 N + 1 } } \sum_{j=-N}^{N} \alpha_{j}(\omega) .
\end{equation}
Equation~(\ref{eq:2_multi_mode_alpha_continuous}) is more involved than Eq.~(\ref{eq:2_multi_mode_alpha_impulse}) due to its dependence on the mode number $j$ in both the denominator and the exponential phase modulation in $\mathcal{E}_{j}$, hence it is difficult to derive a general analytic solution for the frequency response. If we remove the phase modulation, that is we set $m = 0$, then we can derive an expression. As before, we replace the summation over modes with an integral to find the collective field amplitude to be:
\begin{equation}
  \bar{A}(\omega) \simeq \frac{ -i \mathcal{E}_{d} e^{-i \omega t} }{ \delta\omega \left( 2 N + 1 \right) } \mathrm{ln} \left[ \frac{ \kappa - i \left( \omega - \omega_{N} \right) }{ \kappa - i \left( \omega - \omega_{-N} \right) } \right] ,
\end{equation}
where $\omega_{\pm N} = \omega_{c} \pm N \delta\omega$. The collective frequency response is, therefore,
\begin{align}\label{eq:2_multi_mode_response_m0}
  | \bar{A}(\omega) |^{2} &= \frac{ | \mathcal{E}_{d} |^{2} }{ \delta\omega^{2} \left( 2 N + 1 \right)^{2} } \nonumber \\
  &\quad\quad \times \left| \mathrm{ln} \left[ \frac{ \kappa + i \left( \omega - \omega_{N} \right) }{ \kappa + i \left( \omega - \omega_{-N} \right) } \right] \right|^{2} ,
\end{align}
from which we see the effective bandwidth of the multi-mode array filter in the $\omega - \omega_{\pm N}$ terms.

We plot the collective frequency response in Fig.~\ref{fig:2_frequency_response} for three different sizes of phase modulation -- (a) $m = 0$, calculated from Eq.~(\ref{eq:2_multi_mode_response_m0}), (b) $m = 1$ , and (c) $m = 2$ -- for three different effective halfwidths: $N\delta\omega / |\mathcal{E}_{d} = 1$ (blue, solid), 2 (orange, dashed), and 4 (green, dash-dot). By setting $m = 0$, the collective response of the multi-mode array filter produces the ``bunny-ear'' like response, characteristic of the Fourier transform of the positive-sided sinc function in Fig.~\ref{fig:2_sinc_stuff}(d). Applying a slight modulation, with $m = 1$, we see that the collective response closely resembles that of Fig.~\ref{fig:2_sinc_stuff}(f), with much sharper frequency cut-offs than a standard Lorentzian (grey, dotted). We cannot, however, push the phase modulation too far. As previously mentioned, an increase in the size of the modulation also increases the temporal response of the filter. We also see a noticeable dip in the centre of the frequency response when $m = 2$, in Fig.~\ref{fig:2_frequency_response}(c).

These results are promising as they show we can increase the bandwidth of the multi-mode array filter to be almost four times larger than a standard Lorentzian filter, without sacrificing frequency isolation. This is important for two related, yet important reasons: firstly, the wider bandwidth results in a faster temporal response, thus allowing for more accurate photon correlations to be measured; and secondly, the bandwidth of the filter needs to be large enough such that it encompasses the spectral shape of the input field, while still rejecting non-target frequencies. If the filter bandwidth is too narrow compared to the input field -- and thus the lifetime of the cavity is much longer than the lifetime of the source -- the dynamics of the source system are now averaged out over the lifetime of the filter \cite{Nienhuis_1993_EPL_SpectralCorrelationsFluorescence, Nienhuis_1993_PRA_SpectralCorrelationsResonance}.

%==============================================================================%
%                           3 RESONANCE FLUORESCENCE                           %
%==============================================================================%
\section{A Resonantly Driven Two-Level Atom}
\label{sect:3_atom}

Having introduced the multi-mode array filter and investigated its temporal and frequency response to a coherent driving, we now investigate its effectiveness at measuring and calculating frequency-filtered photon correlations. The resonantly driven two-level atom provides an excellent scenario with which to test this due to its relative simplicity and its history of study.

%------------------------------------------------------------------------------%
%-----------------------------------------%
%     3.1 The optical Bloch equations     %
%-----------------------------------------%
\subsection{The optical Bloch equations}
%------------------------------------------------------------------------------%

In this section we will briefly introduce the Mollow triplet and the notation that we will use throughout the rest of this paper. We consider a two-level atom with ground state $\ket{g}$ and excited state $\ket{e}$, with respective eigenfrequencies $\omega_{g}$ and $\omega_{e}$. The atom is coherently driven at the atomic resonance frequency, $\omega_{A} = \omega_{e} - \omega_{g}$, with Rabi frequency $\Omega$, such that the Hamiltonian, in a frame rotating at $\omega_{A}$, is,
\begin{equation}\label{eq:3_2LA_hamiltonian}
  H_{A} = \hbar \frac{ \Omega }{ 2 } \left( \sigma_{+} + \sigma_{-} \right),
\end{equation}
where $\sigma_{+} = \ket{e} \bra{g}$ and $\sigma_{-} = \ket{g} \bra{e}$ are the atomic raising and lowering operators, respectively. To account for spontaneous emission, the Lindblad master equation for the atomic density operator $\rho$ is introduced in the form
\begin{equation}\label{eq:3_2LA_master_equation}
  \ddt[\rho] = \frac{ 1 }{ i \hbar } [ H_{A}, \rho ] + \frac{ \gamma }{ 2 } \Lambda(\sigma_{-})\rho ,
\end{equation}
where $\gamma$ is the atomic excited state decay rate. Expanding the density operator in the atomic state basis, we can rewrite the master equation in terms of the expectation values of the atomic operators:
\begin{equation}
  \ddt \langle \bm{\sigma} \rangle = \bm{M}^{(\bm{\sigma})} \langle \bm{\sigma} \rangle + \bm{B},
\end{equation}
with
\begin{subequations}\label{eq:3_optical_bloch_equations}
  \begin{gather}
    \langle \bm{\sigma} \rangle =
    \begin{pmatrix}
      \langle \sigma_{-} \rangle \\
      \langle \sigma_{+} \rangle \\
      \langle \sigma_{z} \rangle
    \end{pmatrix}, \quad \bm{B} =
    \begin{pmatrix}
      0 \\
      0 \\
      -\gamma
    \end{pmatrix} \\
    \bm{M}^{(\bm{\sigma})} =
    \begin{pmatrix}
      - \frac{\gamma}{2} & 0 & i \frac{\Omega}{2} \\
      0 & -\frac{\gamma}{2} & -i \frac{\Omega}{2} \\
      i \Omega & -i \Omega & -\gamma
    \end{pmatrix} , \label{eq:3_bloch_equations_matrix}
  \end{gather}
\end{subequations}
where $\sigma_{z} = \ket{e}\bra{e} - \ket{g}\bra{g}$. These equations, known as the optical Bloch equations \cite{Bloch_1946_PR_NuclearInduction} provide a concise and simple method for calculating appropriate quantities, such as the population inversion, and will play a vital role in the modelling of the multi-mode-array filtered fluorescence.

\begin{figure}
  \centering
  \includegraphics[width=\linewidth]{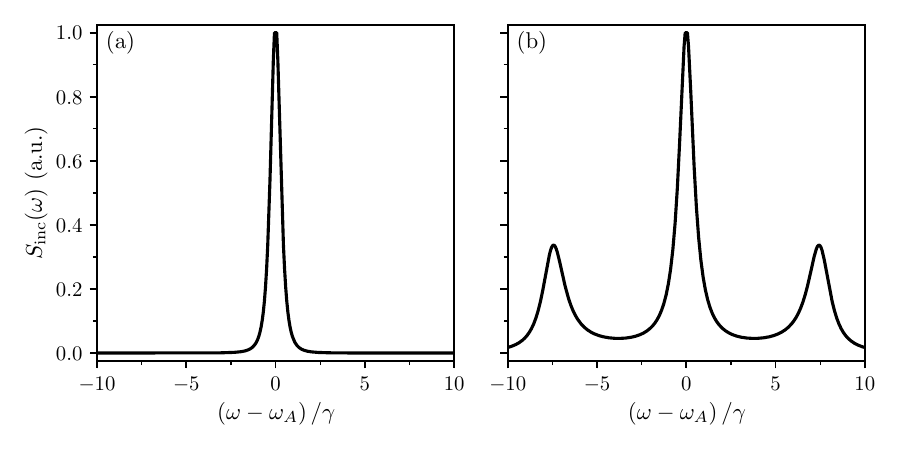}
  \caption{Incoherent power spectrum of the two-level atom for Rabi frequency (a) $\Omega = 0.1 \gamma$ and (b) $\Omega = 7.5 \gamma$.}
  \label{fig:3_2LA_spectrum}
\end{figure}

%------------------------------------------------------------------------------%
%--------------------------------%
%     3.2 The Mollow triplet     %
%--------------------------------%
\subsection{The Mollow triplet}
%------------------------------------------------------------------------------%

The fluorescence spectrum can be calculated from the Fourier transform of the first-order correlation function, using a form of the Wiener-Khinchin theorem \cite{Wiener_1930_AM_GeneralizedHarmonicAnalysis, Khinchin_1934_MA_KorrelationstheorieDerStationaren}:
\begin{equation}
  S(\omega) = \frac{1}{2 \pi} \int_{-\infty}^{\infty} e^{i \omega \tau} g^{(1)}_{ss}(\tau) \mathrm{d} \tau,
\end{equation}
where
\begin{align}
  g^{(1)}_{ss}(\tau) &= \lim_{t \rightarrow \infty} \frac{ \langle \sigma_{+}(t) \sigma_{-}(t + \tau) \rangle }{ \sqrt{ \langle \sigma_{+} \sigma_{-}(t) \rangle \langle \sigma_{+} \sigma_{-} (t + \tau) \rangle } } \nonumber \\
  &= \frac{ \langle \sigma_{+}(0) \sigma_{-}(\tau) \rangle_{ss} }{ \langle \sigma_{+} \sigma_{-} \rangle_{ss} }
\end{align}
is the normalised steady-state first-order correlation function for the driven two-level atom. In the steady state there exist quantum fluctuations, describable with operators $\Delta \sigma_{\pm} = \sigma_{\pm} - \langle \sigma_{\pm} \rangle_{ss}$, and so we decompose the fluorescence spectrum into coherent and incoherent components:
\begin{subequations}
  \begin{align}
    S_{\mathrm{coh}}(\omega) &= \frac{1}{2 \pi} \int_{-\infty}^{\infty} e^{i \omega \tau} \frac{\langle \sigma_{+} \rangle_{ss} \langle \sigma_{-} \rangle_{ss}}{\langle \sigma_{+} \sigma_{-} \rangle_{ss}} \mathrm{d}\tau, \\
    %-----------------
    S_{\mathrm{inc}}(\omega) &= \frac{1}{2 \pi} \int_{-\infty}^{\infty} e^{i \omega \tau} \frac{\langle \Delta \sigma_{+}(0) \Delta \sigma_{-}(\tau) \rangle_{ss}}{\langle \sigma_{+} \sigma_{-} \rangle_{ss}} \mathrm{d}\tau .
  \end{align}
\end{subequations}
In the weak driving regime, $\Omega \ll \gamma$, the incoherent power spectrum of the atom is Lorentzian, centred at the atomic frequency $\omega_{A}$, and with halfwidth $\gamma / 2$. As the Rabi frequency increases, and thus incoherent scattering dominates, side-peaks emerge at $\omega_{A} \pm \Omega$, giving shape to the Mollow triplet \cite{Mollow_1975_PRA_PureStateAnalysis, Mollow_1975_PRA_PureStateAnalysis}, as shown in Fig.~\ref{fig:3_2LA_spectrum}.

%------------------------------------------------------------------------------%
%-----------------------------------%
%     3.3 Atomic dressed states     %
%-----------------------------------%
\subsection{Atomic dressed states}
%------------------------------------------------------------------------------%

\begin{figure}
  \centering
  \includegraphics[width=\linewidth]{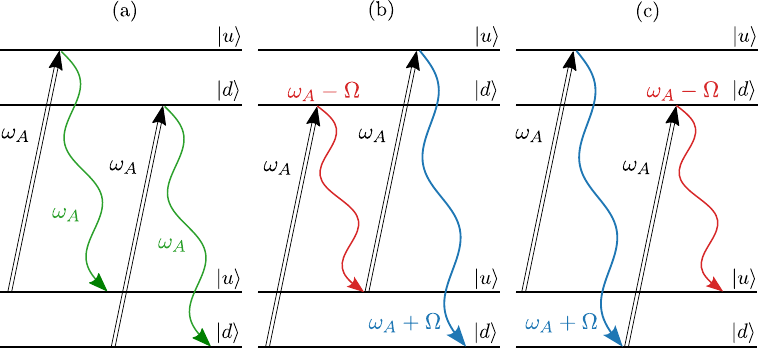}
  \caption{The three components of the Mollow triplet arise from photon emissions down the ladder of atomic dressed states.}
  \label{fig:3_2LA_dressed_states}
\end{figure}

Each of the Mollow triplet peaks can be attributed to transitions amongst the atomic dressed states \cite{Schrama_1992_PRA_IntensityCorrelationsComponents, CohenTannoudji_1979_PTotRSoLSAMaPS_AtomsStrongLight} -- eigenstates of the combined atomic and quantised field Hamiltonians. By diagonalising the Hamiltonian, Eq.~(\ref{eq:3_2LA_hamiltonian}), we can express the eigenstates in terms of the atomic bare states as:
\begin{subequations}\label{eq:3_2LA_dressed_states}
  \begin{align}
    H_{A} \ket{u} = + \hbar \frac{ \Omega }{ 2 } \ket{u} , \quad \ket{u} = \frac{ 1 }{ \sqrt{2} } \Big( \ket{e} + \ket{g} \Big) , \\
    %-----------------
    H_{A} \ket{d} = - \hbar \frac{ \Omega }{ 2 } \ket{d} , \quad \ket{d} = \frac{ 1 }{ \sqrt{2} } \Big( \ket{e} - \ket{g} \Big) .
  \end{align}
\end{subequations}
These dressed states form an energy level ladder, where each pair are separated by an energy of $\hbar \omega_{A}$ in the quantised radiation field. In the strong driving regime, where the two dressed states are well separated, three sets of dressed state transitions occur, as depicted in Fig.~\ref{fig:3_2LA_dressed_states}: $\ket{d} \rightarrow \ket{u}$ with frequency $\omega_{A} - \Omega$, $\ket{u} \rightarrow \ket{u}$ and $\ket{d} \rightarrow \ket{d}$ with frequency $\omega_{A}$, and $\ket{u} \rightarrow \ket{d}$ with frequency $\omega_{A} + \Omega$. We can express the atomic operators in terms of these dressed states, allowing us to rewrite the atomic master equation in the dressed state basis. Transforming into a rotating frame with unitary evolution operator $\mathcal{U}(t) = \exp\left[ H_{A} t / i \hbar \right]$, Eq.~(\ref{eq:3_2LA_master_equation}) becomes
\begin{equation}\label{eq:3_atom_master_equation_interaction}
  \ddt[\rho] = \frac{ \gamma }{ 2 } \Lambda \left( \tilde{\sigma}_{-}(t) \right) \tilde{\rho}(t) ,
\end{equation}
with $\tilde{\rho}(t) = \mathcal{U}^{\dagger}(t) \rho~ \mathcal{U}(t)$ and
\begin{align}
  \tilde{\sigma}_{\pm}(t) &= \mathcal{U}^{\dagger}(t) \sigma_{\pm} \mathcal{U}(t) \nonumber \\
  &= \frac{ 1 }{ 2 } \left[ \sigma_{z}^{D} \pm \left( \sigma_{-}^{D} e^{-i \Omega t} - \sigma_{+}^{D} e^{i \Omega t} \right) \right],
\end{align}
where we have introduced the \textit{dressed-state operators}:
\begin{subequations}
  \begin{gather}
    \sigma^{D}_{-} \equiv \ket{d} \bra{u}, \quad \sigma^{D}_{+} \equiv \ket{u} \bra{d}, \\
    %-----------------
    \sigma^{D}_{z} \equiv \ket{u} \bra{u} - \ket{d} \bra{d}.
  \end{gather}
\end{subequations}
Using these operators in Eq.~(\ref{eq:3_atom_master_equation_interaction}) and making the secular approximation by dropping any rapidly oscillating terms (assuming $\Omega \gg \gamma$), the master equation reduces to \cite{Nienhuis_1993_PRA_SpectralCorrelationsResonance}:
\begin{align}\label{eq:3_dressed_state_master_equation}
  \ddt[\rho] = -i \frac{ \Omega }{ 2 } [ \sigma_{z}^{D} , \rho ] &+ \frac{ \gamma }{ 8 } \Lambda \left( \sigma_{-}^{D} \right) \rho + \frac{ \gamma }{ 8 } \Lambda \left( \sigma_{+}^{D} \right) \rho \nonumber \\
  &+ \frac{ \gamma }{ 4 } \Lambda \left( \sigma_{z}^{D} \right) \rho.
\end{align}
From this equation we can identify the three processes corresponding to the three Mollow triplet peaks: a $\ket{u} \rightarrow \ket{d}$ transition with operator $\sigma_{-}^{D}$, a $\ket{d} \rightarrow \ket{u}$ transition with operator $\sigma_{+}^{D}$, and a dephasing term with operator $\sigma_{z}^{D}$, corresponding to the $\ket{u} \rightarrow \ket{u}$ and $\ket{d} \rightarrow \ket{d}$ transitions.

%------------------------------------------------------------------------------%
%------------------------------------------%
%     3.3.1 Auto-correlation functions     %
%------------------------------------------%
\subsubsection{Auto-correlation functions}
%------------------------------------------------------------------------------%

From Eq.~(\ref{eq:3_dressed_state_master_equation}) we can derive a set of simple decoupled moment equations similar to the optical Bloch equations, Eqs.~(\ref{eq:3_optical_bloch_equations}),
\begin{subequations}
  \begin{align}
    \ddt \langle \sigma_{-}^{D} \rangle &= -\left( \frac{ 3 \gamma }{ 4 } + i \Omega \right) \langle \sigma_{-}^{D} \rangle , \\
    %-----------------
    \ddt \langle \sigma_{+}^{D} \rangle &= -\left( \frac{ 3 \gamma }{ 4 } - i \Omega \right) \langle \sigma_{+}^{D} \rangle , \\
    %-----------------
    \ddt \langle \sigma_{z}^{D} \rangle &= -\frac{ \gamma }{ 2 } \langle \sigma_{z}^{D} \rangle .
  \end{align}
\end{subequations}
We can then use the quantum regression equations (see Refs.~\cite{Lax_1963_PR_FormalTheoryQuantum, Lax_1967_PR_QuantumNoiseX} and Sec.~1.5 of \cite{Carmichael_StatisticalMethodsQuantum1} for details) to derive second-order correlation functions for the central-, left-, and right-peak photon transitions, respectively:
\begin{subequations}\label{eq:3_dressed_state_auto_correlation_eqns}
  \begin{align}
    g^{(2)}_{C}(\tau) &= \frac{ \langle \sigma_{z}^{D}(0) \sigma_{z}^{D} \sigma_{z}^{D}(\tau) \sigma_{z}^{D}(0) \rangle_{ss} }{ \langle \sigma_{z}^{D} \sigma_{z}^{D} \rangle_{ss}^{2} } = 1 , \label{eq:3_g2_dressed_C} \\
    %-----------------
    g^{(2)}_{L}(\tau) &= \frac{ \langle \sigma_{-}^{D}(0) \sigma_{-}^{D} \sigma_{+}^{D}(\tau) \sigma_{+}^{D}(0) \rangle_{ss} }{ \langle \sigma_{-}^{D} \sigma_{+}^{D} \rangle_{ss}^{2} } = 1 - e^{-\frac{ \gamma }{ 2 } \tau } , \label{eq:3_g2_dressed_L} \\
    %-----------------
    g^{(2)}_{R}(\tau) &= \frac{ \langle \sigma_{+}^{D}(0) \sigma_{+}^{D} \sigma_{-}^{D}(\tau) \sigma_{-}^{D}(0) \rangle_{ss} }{ \langle \sigma_{+}^{D} \sigma_{-}^{D} \rangle_{ss}^{2} } = 1 - e^{-\frac{ \gamma }{ 2 } \tau } . \label{eq:3_g2_dressed_R}
  \end{align}
\end{subequations}
The left- and right-peak photons both exhibit antibunching characteristic of the two-level atom, as expected from the de-excitation paths illustrated in Fig.~\ref{fig:3_2LA_dressed_states}(b) and (c). Central-peak photons, however, are entirely second-order coherent, as the dressed-state density matrix remains unchanged after an emission.

%------------------------------------------------------------------------------%
%-------------------------------------------%
%     3.3.2 Cross-correlation functions     %
%-------------------------------------------%
\subsubsection{Cross-correlation functions}
%------------------------------------------------------------------------------%

We can also derive expressions for cross-correlation functions of the dressed state transitions, allowing us to look at two separate photon emissions:
\begin{subequations}\label{eq:3_dressed_state_cross_corr}
  \begin{align}
    g^{(2)}_{RC}(\tau) &= \frac{ \langle \sigma_{-}^{D}(0) \sigma_{z}^{D} \sigma_{z}^{D}(\tau) \sigma_{+}^{D}(0) \rangle_{ss} }{ \langle \sigma_{-}^{D} \sigma_{+}^{D} \rangle_{ss} \langle \sigma_{z}^{D} \sigma_{z}^{D} \rangle_{ss} } = 1 , \label{eq:3_dressed_state_cross_corr_RC}  \\
    %-----------------
    g^{(2)}_{RL}(\tau) &= \frac{ \langle \sigma_{-}^{D}(0) \sigma_{+}^{D} \sigma_{-}^{D}(\tau) \sigma_{+}^{D}(0) \rangle_{ss} }{ \langle \sigma_{+}^{D} \sigma_{-}^{D} \rangle_{ss} \langle \sigma_{-}^{D} \sigma_{+}^{D} \rangle_{ss} } \nonumber \\
    &= 1 + e^{-\frac{ \gamma }{ 2 } \tau } . \label{eq:3_dressed_state_cross_corr_RL}
  \end{align}
\end{subequations}
Given an initial detection of a right-peak photon, Eq.~(\ref{eq:3_dressed_state_cross_corr_RL}) shows that cross-correlations between the two side-peaks are strong, as indicated in the transition order in Figs.~\ref{fig:3_2LA_dressed_states}(b) and \ref{fig:3_2LA_dressed_states}(c), while correlations between the central peak and either of the side-peaks remain second-order coherent. When considering frequency filtered correlations, however, these expressions only hold for \textit{long-time behaviour}, where $\tau \gg \gamma^{-1}$. In the short-time behaviour, $\tau \ll \gamma^{-1}$, there are more complicated dynamics occurring. For example, for an emission of a right-peak photon and a central-peak photon, the emission can occur via two different cascade pathways, $\ket{u} \rightarrow \ket{d} \rightarrow \ket{d}$ and $\ket{u} \rightarrow \ket{u} \rightarrow \ket{d}$. While these two pathways technically describe two different de-excitations path, they both correspond to an emission starting in the upper state $\ket{u}$ and ending in the lower state $\ket{d}$. On a short time scale, there is destructive interference between the two different time orderings. As derived by Schrama et al. \cite{Schrama_1992_PRA_IntensityCorrelationsComponents}, the frequency-filtered photon correlations for the ideally separated Mollow triplet are:
\begin{subequations}\label{eq:3_schrama_cross_eqns}
  \begin{gather}
    g^{(2)}_{RC}(\tau) = \left( 1 - e^{-K \tau} \right) , \label{eq:3_g2_dressed_RC} \\
    g^{(2)}_{RL}(\tau) = e^{-\frac{ \gamma }{ 2 } \tau} - 1 + \frac{ 1 }{ 2 } \left( 2 - e^{-K \tau} \right)^{2} + \frac{ 1 }{ 2 } e^{-2 K \tau} ,  \label{eq:3_g2_dressed_RL}
  \end{gather}
\end{subequations}
with $K$ given by Eq.~(\ref{eq:2_multi_mode_array_bandwidths}).

%==============================================================================%
%                      4 FREQUENCY FILTERED MOLLOW TRIPLET                     %
%==============================================================================%
\section{The Frequency-Filtered Mollow Triplet}
\label{sect:4_moments_g1}

Having introduced the multi-mode array filter in Section~\ref{sect:2_filtering}, and our source system in Section~\ref{sect:3_atom}, we now merge the two systems together. To model the frequency filtering, we treat the two-level as a source subsystem, and cascade the output fluorescence into the multi-mode array filter using \textit{cascaded open systems theory} \cite{Carmichael_1993_PRL_QuantumTrajectoryTheory, Gardiner_1993_PRL_DrivingQuantumSystem}. The Hamiltonian for this cascaded system is then
\begin{align}\label{eq:4_hamiltonian}
  H = H_{A} &+ \sum_{j=-N}^{N} \hbar \Delta\omega_{j} a_{j}^{\dagger} a_{j} \\
  &\quad + \frac{ i \hbar }{ 2 } \sum_{j=-N}^{N} \left( \mathcal{E}_{j}^{*} a_{j} \sigma_{+} - \mathcal{E}_{j} a_{j}^{\dagger} \sigma_{-} \right) ,
\end{align}
with atomic Hamiltonian, Eq.~(\ref{eq:3_2LA_hamiltonian}), frequency detuning of the $j^{\mathrm{th}}$ mode from atomic resonance, $\Delta\omega_{j} = \left( \omega_{0} + j \delta\omega \right) - \omega_{A}$, and mode-dependent coupling,
\begin{equation}
  \mathcal{E}_{j} = \sqrt{ \frac{ \gamma \kappa }{ 2 N + 1 } } e^{i m j \pi / N} . \label{eq:4_Ej_one_filter}
\end{equation}
The atomic fluorescence is evenly cascaded into an array of $2N + 1$ tunable single-mode cavities, as depicted in Fig.~\ref{fig:4_filter_schematic_and_moment_flowchart}(a). Each mode is modelled as a ring cavity, with two perfect mirrors and two lossy mirrors, each with a loss rate of $\kappa / 2$. This results in two separate output channels: a reflection channel consisting of the reflection of photons from, and the transmission of photons out through the input mirror; and a transmission channel consisting of photons emitted out through the second lossy mirror. Each channel is directed towards a separate photodetector (see Fig.~\ref{fig:4_filter_schematic_and_moment_flowchart}(b)). As noted earlier, all input channels to the beam splitters and cavity coupling mirrors that do not emanate from the source are assumed to be vacuum, and thus make no contribution to the detection process. The total cascaded system master equation is then \cite{Gardiner_1993_PRL_DrivingQuantumSystem, Carmichael_1993_PRL_QuantumTrajectoryTheory}
\begin{equation}\label{eq:4_master_equation}
  \ddt[\rho] = \frac{ 1 }{ i \hbar } [ H , \rho ] + \frac{ \kappa }{ 2 } \sum_{j=-N}^{N} \Lambda \left( a_{j} \right) \rho + \frac{ 1 }{ 2 } \sum_{j=-N}^{N} \Lambda \left( C_{j} \right) \rho,
\end{equation}
where
\begin{equation}
  C_{j} = \sqrt{ \frac{ \gamma }{ 2 N + 1 } } e^{i m j \pi / N} \sigma_{-} + \sqrt{ \kappa } a_{j},
\end{equation}
is the cascaded decay operator, corresponding to the reflection channel.

%------------------------------------------------------------------------------%
%--------------------------------------%
%     4.1 Moment equation approach     %
%--------------------------------------%
\subsection{Moment equation approach}
%------------------------------------------------------------------------------%

\begin{figure*}[t]
  \centering
  \includegraphics[width=\linewidth]{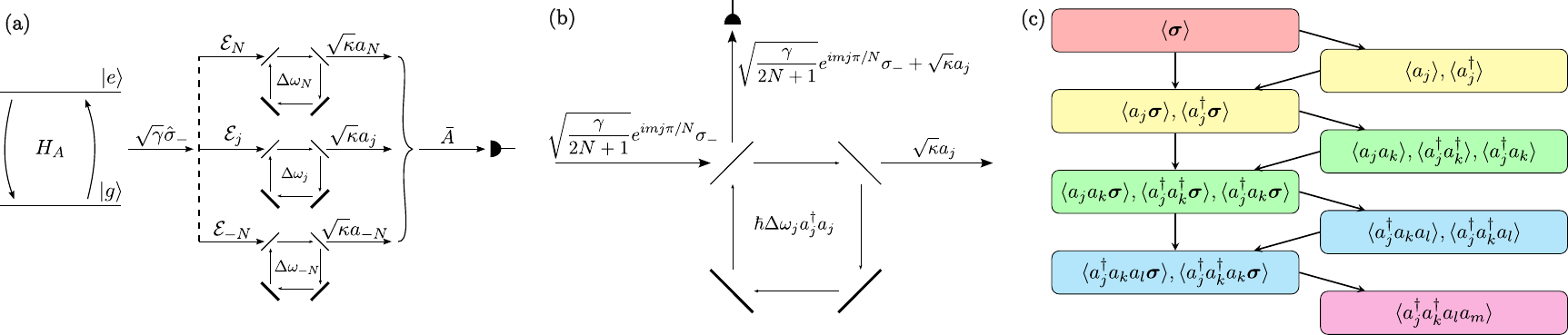}
  \caption{(a) Full schematic of the multi-mode array-filtered two-level atom, with (b) detail of the decay channels of the $j^{\mathrm{th}}$ mode. The phase-shifted fluorescence is evenly cascaded into a ring cavity with two perfectly reflecting mirrors (bold). The reflection (and transmission) from the input mirror and transmission from the output mirror are sent to separate photo-detectors. (c) The atomic and filter-mode operator moment equations couple in a cascaded scheme, where lower-order moments can be solved independently of higher-order moments. The colour of the boxes indicate the order of filter-mode operators, and thus the order in which they can be computed, with the arrows indicating the coupling of lower-order to higher-order moments.}
  \label{fig:4_filter_schematic_and_moment_flowchart}
\end{figure*}

With the master equation, Eq.~(\ref{eq:4_master_equation}), we can solve for the multi-mode array-filtered first- and second-order correlation functions using the quantum regression equations \cite{Lax_1963_PR_FormalTheoryQuantum, Lax_1967_PR_QuantumNoiseX}. For this filtering method, we allow for a large number of cavity modes, $N \gg 1$, each with an infinite dimension Hilbert space. Even by taking a conservative photon number truncation of up to two photons, the size of the total system's density operator grows as $\left( 2 \times 3^{2 N + 1} \right)^{2}$. The computations then prove to be extremely demanding, and as a result we were limited to $N = 20$ on our current hardware.

Fortunately, we can instead find exact solutions of the two-time correlation functions by calculating the \textit{moment equations}. For some operator $X$, we can derive the rate equation of the expectation value by using the master equation, with
\begin{equation}
  \ddt \langle X \rangle = \ddt \mathrm{tr} \left\{ X \rho \right\} = \mathrm{tr} \left\{ X \frac{ \mathrm{d} \rho }{ \mathrm{d} t } \right\} .
\end{equation}
As an example, we find that the second-order moment $\langle a_{j}^{\dagger} a_{k} \rangle$ evolves as
\begin{align}
  \ddt \langle a_{j}^{\dagger} a_{k} \rangle &= \mathrm{tr} \left\{ a_{j}^{\dagger} a_{k} \ddt[\rho] \right\} \nonumber \\
  &= -\left[ 2 \kappa - i \left( \Delta\omega_{j} - \Delta\omega_{k} \right) \right] \langle a_{j}^{\dagger} a_{k} \rangle \nonumber \\
  &\quad\quad - \mathcal{E}_{j}^{*} \langle a_{k} \sigma_{+} \rangle - \mathcal{E}_{j} \langle a_{j}^{\dagger} \sigma_{-} \rangle .
\end{align}
This operator moment is dependent only on itself, and two ``lower order'' moments, $\langle a_{k} \sigma_{+} \rangle$ and $\langle a_{j}^{\dagger} \sigma_{-} \rangle$. These moments then evolve with equations similar to optical Bloch equations:
\begin{subequations}\label{eq:4_aj_sigma_eqns}
  \begin{align}
    \ddt \langle a_{j} \sigma_{-} \rangle &= -\left( \frac{ \gamma }{ 2 } + \kappa + i \Delta\omega_{j} \right) \langle a_{j} \sigma_{-} \rangle \nonumber \\
    &\quad + i \frac{ \Omega }{ 2 } \langle a_{j} \sigma_{z} \rangle , \\
    \ddt \langle a_{j} \sigma_{+} \rangle &= -\left( \frac{ \gamma }{ 2 } + \kappa + i \Delta\omega_{j} \right) \langle a_{j} \sigma_{+} \rangle \nonumber \\
    &\quad - i \frac{ \Omega }{ 2 } \langle a_{j} \sigma_{z} \rangle , \\
    \ddt \langle a_{j} \sigma_{z} \rangle &= i \Omega \langle a_{j} \sigma_{-} \rangle - i \Omega \langle a_{j} \sigma_{+} \rangle \nonumber \\
    &\quad - \left( \gamma + \kappa + i \Delta\omega_{j} \right) \langle a_{j} \sigma_{z} \rangle  .
  \end{align}
\end{subequations}
We can write Eqs.~(\ref{eq:4_aj_sigma_eqns}) in matrix form
\begin{align}
  \ddt
  \begin{pmatrix}
    a_{j} \sigma_{-} \\
    a_{j} \sigma_{+} \\
    a_{j} \sigma_{z}
  \end{pmatrix}
  &= \left[ \bm{M}^{(\bm{\sigma})} - \left( \kappa + i \Delta\omega_{j} \right) \mathds{1} \right]
  \begin{pmatrix}
    a_{j} \sigma_{-} \\
    a_{j} \sigma_{+} \\
    a_{j} \sigma_{z}
  \end{pmatrix}
  \nonumber \\
  &\quad +
  \begin{pmatrix}
    0 \\
    -\frac{ 1 }{ 2 } \mathcal{E}_{j} \left( \langle \sigma_{z} \rangle + 1 \right) \\
    -\gamma \langle a_{j} \rangle + \mathcal{E}_{j} \langle \sigma_{-} \rangle
  \end{pmatrix} ,
\end{align}
where the evolution matrix $\bm{M}^{(\bm{\sigma})}$ is given by Eq.~(\ref{eq:3_bloch_equations_matrix}). Introducing the operator vector notation,
\begin{equation}
  \langle X \bm{\sigma} \rangle =
  \begin{pmatrix}
    \langle X \sigma_{-} \rangle \\
    \langle X \sigma_{+} \rangle \\
    \langle X \sigma_{z} \rangle
  \end{pmatrix} ,
\end{equation}
the first six coupled moment equations are:
\begin{subequations}\label{eq:4_first_few_moment_equations}
  \begin{align}
    \ddt \langle \bm{\sigma} \rangle &= \bm{M}^{(\bm{\sigma})} \langle \bm{\sigma} \rangle +
    \begin{pmatrix}
      0 \\
      0 \\
      -\gamma
    \end{pmatrix} , \label{eq:4_moments_eqns_sig} \\
    %-----------------
    \ddt \langle a_{j} \rangle &= -\left( \kappa + i \Delta\omega_{j} \right) \langle a_{j} \rangle - \mathcal{E}_{j} \langle \sigma_{-} \rangle , \\
    %-----------------
    \ddt \langle a_{j}^{\dagger} \rangle &= -\left( \kappa - i \Delta\omega_{j} \right) \langle a_{j}^{\dagger} \rangle - \mathcal{E}_{j}^{*} \langle \sigma_{+} \rangle , \label{eq:4_moments_eqns_at} \\
    %-----------------
    \ddt \langle a_{j} \bm{\sigma} \rangle &= \left[ \bm{M}^{(\bm{\sigma})} - \left( \kappa + i \Delta\omega_{j} \right) \mathds{1} \right] \langle a_{j} \bm{\sigma} \rangle \nonumber \\
    &\quad +
    \begin{pmatrix}
      0 \\
      -\frac{ 1 }{ 2 } \mathcal{E}_{j} \left( \langle \sigma_{z} \rangle + 1 \right) \\
      -\gamma \langle a_{j} \rangle + \mathcal{E}_{j} \langle \sigma_{-} \rangle
    \end{pmatrix} , \\
    %-----------------
    \ddt \langle a_{j}^{\dagger} \bm{\sigma} \rangle &= \left[ \bm{M}^{(\bm{\sigma})} - \left( \kappa - i \Delta\omega_{j} \right) \mathds{1} \right] \langle a_{j}^{\dagger} \bm{\sigma} \rangle \nonumber \\
    &\quad +
    \begin{pmatrix}
      -\frac{ 1 }{ 2 } \mathcal{E}_{j}^{*} \left( \langle \sigma_{z} \rangle + 1 \right) \\
      0 \\
      -\gamma \langle a_{j}^{\dagger} \rangle + \mathcal{E}_{j}^{*} \langle \sigma_{+} \rangle
    \end{pmatrix} , \\
    %-----------------
    \ddt \langle a_{j}^{\dagger} a_{k} \rangle &= -\left[ 2 \kappa - i \left( \Delta\omega_{j} - \Delta\omega_{k} \right) \right] \langle a_{j}^{\dagger} a_{k} \rangle \nonumber \\
    &\quad - \mathcal{E}_{j}^{*} \langle a_{k} \sigma_{+} \rangle - \mathcal{E}_{j} \langle a_{j}^{\dagger} \sigma_{-} \rangle .
  \end{align}
\end{subequations}

\begin{figure*}[t]
  \centering
  \includegraphics[width=\linewidth]{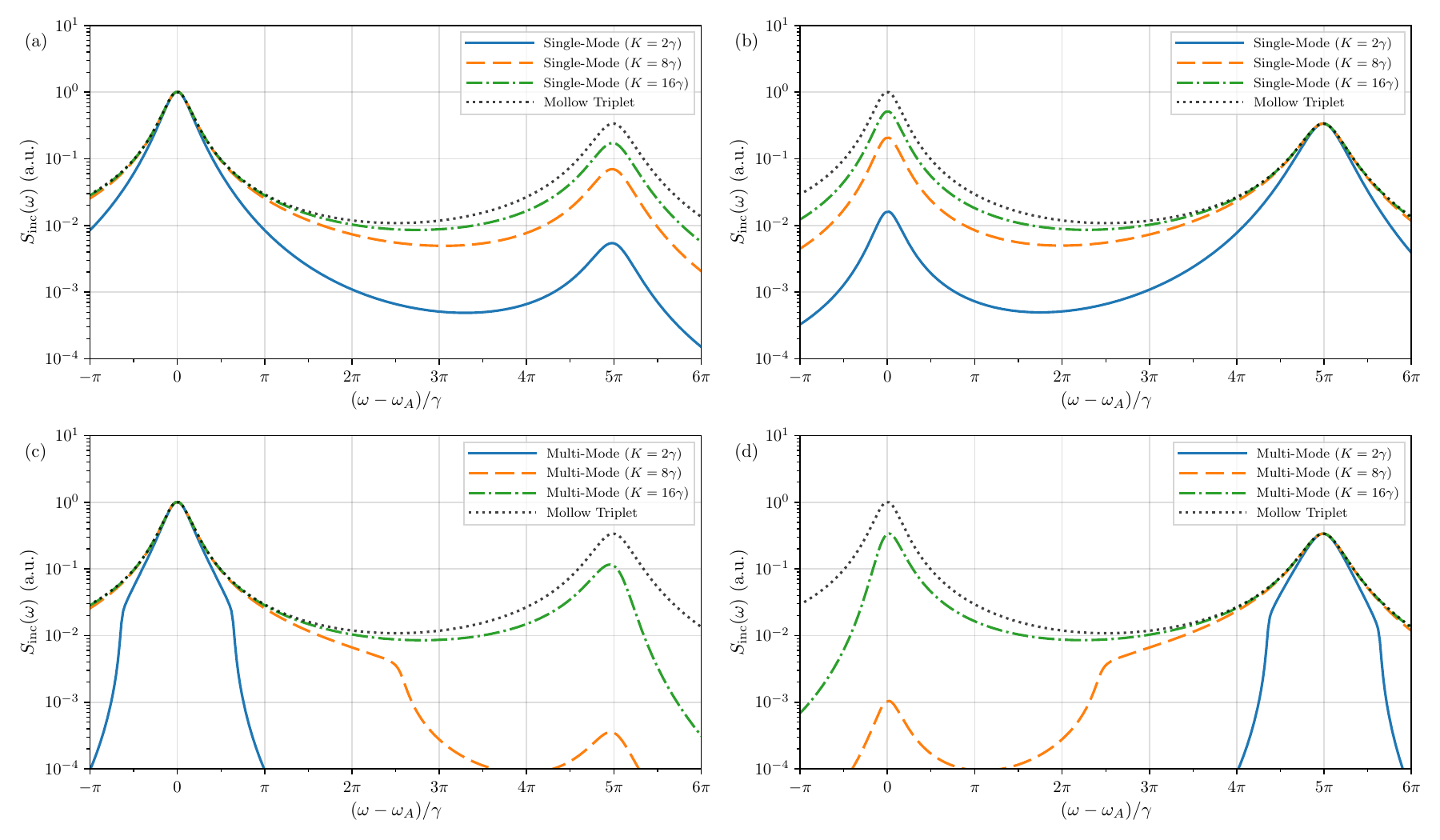}
  \caption{Frequency-filtered incoherent power spectrum of a single-mode (a, b), $N = 0$, and multi-mode (c, d), $N = 80$, filter. The filter resonance is resonant with the (a, c) central-peak, $\Delta\omega_{0} = 0$,  and the (b, d) right-peak, $\Delta\omega_{0} = 5 \pi \gamma$, of the Mollow triplet, with halfwidths $K = 2 \gamma$ (blue, solid), and $K = 4 \gamma$ (orange, dashed), $K = 16 \gamma$ (green, dot-dash). The filtered spectra are compared against the unfiltered Mollow triplet (grey, dotted). Other parameters are $\Omega = 5 \pi \gamma, m = 1$ and, for the multi-mode array filter $\delta\omega = K / N, \kappa = 2.5 \delta\omega$.}
  \label{fig:4_filtered_spectrum}
\end{figure*}

A coherent coupling, like that of the Jaynes-Cummings model, would introduce a bidirectional coupling, with lower-order terms depending on higher-order terms. Hence, just like the Jaynes-Cummings model, it would be impossible to derive a closed set of equations for the second-order correlation function. The unidirectional cascading, however, enforces a simple structure to the coupled moment equations: higher-order moments are dependent only on lower-order terms, as seen in Eqs.~\ref{eq:4_first_few_moment_equations}. We depict the couplings in order of calculations in Fig.~\ref{fig:4_filter_schematic_and_moment_flowchart}(c), up to the fourth-order moment $\langle a_{j}^{\dagger} a_{k}^{\dagger} a_{l} a_{m} \rangle$ needed to calculate the second-order correlation function. With zero back-action of the filter onto the atom, the optical Bloch equations serve as the source of all of the dynamics of the system, and can be easily solved as a $3 \times 3$ matrix of coupled equations.

%------------------------------------------------------------------------------%
%----------------------------------------------%
%     4.2 First-order correlation function     %
%----------------------------------------------%
\subsection{First-order correlation function}
%------------------------------------------------------------------------------%

The normalised first-order correlation function for the multi-mode array filter in the steady-state limit is
\begin{equation}\label{eq:4_filtered_g1}
  g^{(1)}_{\mathrm{filter}} = \frac{ \langle A^{\dagger}(\tau) A(0) \rangle_{ss} }{ \langle A^{\dagger} A \rangle_{ss} },
\end{equation}
where we have introduced the \textit{non-normalised} collective mode annihilation operator, $ A = \sqrt{2N + 1 } \bar{A}$. Decomposing the filtered power spectrum into coherent and incoherent scattering components, the \textit{frequency-filtered incoherent power spectrum} is then given by:
\begin{equation}
  S_{\mathrm{inc}}(\omega) = \frac{ 1 }{ 2 \pi } \int_{-\infty}^{\infty} e^{ i \omega \tau } \frac{ \langle \Delta A^{\dagger}(\tau) \Delta A(0) \rangle }{ \langle A^{\dagger} A \rangle_{ss} } \mathrm{d}\tau .
\end{equation}
We can derive an analytic expression for the time dependence of the first-order correlation function with only two sets of equations, Eqs.~(\ref{eq:4_moments_eqns_sig}) and (\ref{eq:4_moments_eqns_at}). For the initial conditions, however, we need to solve up to the second-order moment $\langle a_{j}^{\dagger} a_{k} \rangle_{ss}$. While these can also be solved analytically, it is a tedious exercise and so we calculate these numerically. The \textit{first-order correlation function for the multi-mode array-filtered two-level atom} is then:
\begin{align}\label{eq:4_filtered_G1}
  \langle \Delta A^{\dagger}(\tau) \Delta A(0) \rangle &= C_{1} e^{- \frac{ \gamma }{ 2 } \tau } + C_{2} e^{- \left( \frac{ 3 \gamma }{ 4 } - \delta \right) \tau } \nonumber \\
  &\quad + C_{3} e^{- \left( \frac{ 3 \gamma }{ 4 } + \delta \right) \tau } + C_{4} e^{- \kappa \tau } ,
\end{align}
where
\begin{subequations}\label{eq:4_g1_coefficients}
  \begin{align}
    %-----------------
    C_{1} &= \sum_{j=-N}^{N} \frac{ \mathcal{E}_{j}^{*} }{ 2 } \frac{ \Delta_{-} + \Delta_{+}  }{ \gamma / 2 - \kappa + i \Delta\omega_{j} } , \\
    %-----------------
    C_{2} &= \sum_{j=-N}^{N} \frac{ -\mathcal{E}_{j}^{*} }{ 4 \delta } \frac{ \left( \gamma / 4 + \delta \right) \left( \Delta_{-} - \Delta_{+} \right) + i \Omega \Delta_{z} }{ 3 \gamma / 4 - \delta - \kappa + i \Delta\omega_{j} } , \\
    %-----------------
    C_{3} &= \sum_{j=-N}^{N} \frac{ \mathcal{E}_{j}^{*} }{ 4 \delta } \frac{ \left( \gamma / 4 - \delta \right) \left( \Delta_{-} - \Delta_{+} \right) + i \Omega \Delta_{z} }{ 3 \gamma / 4 + \delta - \kappa + i \Delta\omega_{j} } , \\
    %-----------------
    C_{4} &= \sum_{j=-N}^{N} \left( e^{i \Delta\omega_{j} \tau } \langle \Delta a_{j}^{\dagger} \Delta A \rangle_{ss} \right) \nonumber \\
    &\quad - C_{1} - C_{2} - C_{3} ,
  \end{align}
\end{subequations}
with
\begin{subequations}\label{eq:4_g1_coefficients_stuff}
  \begin{align}
    \delta &= \sqrt{ \left( \frac{ \gamma }{ 4 } \right)^{2} - \Omega^{2} } , \\
    %-----------------
    \Delta_{-} &= \langle \Delta \sigma_{-} \Delta A \rangle_{ss} , \\
    %-----------------
    \Delta_{+} &= \langle \Delta \sigma_{+} \Delta A \rangle_{ss} , \\
    %-----------------
    \Delta_{z} &= \langle \Delta \sigma_{z} \Delta A \rangle_{ss} .
  \end{align}
\end{subequations}
While we do not give explicit expressions for the steady state expectation values in Eqs.~(\ref{eq:4_g1_coefficients}) and (\ref{eq:4_g1_coefficients_stuff}), we can compute these numerically from the moment equations given in the Appendix.

Using Eq.~(\ref{eq:4_filtered_G1}), we compute the incoherent power spectrum of the single- and multi-mode array filters for increasing halfwidths, shown in Fig.~\ref{fig:4_filtered_spectrum}. For the multi-mode array filter, we choose a large number of modes, $N = 80$, to ensure a sharp cut-off in the frequency response. Comparing the single-mode filter, Figs.~\ref{fig:4_filtered_spectrum}(a) and \ref{fig:4_filtered_spectrum}(b), to the unfiltered Mollow triplet (grey dotted line), we see two distinct features. Firstly, for the smallest halfwidth shown, $K = 2 \gamma$, the output resonance is much narrower than the target peak. Increasing the halfwidth worsens the frequency isolation, as the Lorentzian response overlaps with the other peaks of the Mollow triplet.

Figures~\ref{fig:4_filtered_spectrum}(c) and \ref{fig:4_filtered_spectrum}(d) show a dramatic drop-off in the incoherent power spectra of the multi-mode array filtered fluorescence. As with the single-mode filter, for the smallest halfwidth, $K = 2 \gamma$, the linewidth of the filtered fluorescence is much narrower than the natural linewidth. With the largest halfwidth, $K = 16 \gamma$, the multi-mode array filter performs similarly to the single-mode filter, as the halfwidth encompasses both side-peaks of the Mollow triplet. The intermediate halfwidth,  at $K = 8 \gamma$, however, demonstrates the sharp frequency response cut-off of the multi-mode array filter. This gives a good indication of the optimal halfwidth regime for accurately measured frequency-filtered photon correlations.

%==============================================================================%
%                   5 FREQUENCY-FILTERED PHOTON-CORRELATIONS                   %
%==============================================================================%
\section{Frequency-Filtered Photon Correlations}
\label{sect:6_photon_correlations}

To measure and calculate frequency-filtered photon correlations, we propose a model where the two-level atom is simultaneously coupled into \textit{two} separate multi-mode array filters, which we label as filters \textit{A} and \textit{B}. For simplicity's sake, we assume these two filters to have the same number of modes, frequency spacing, and mode halfwidth -- $N, \delta\omega$, and $\kappa$ -- but with different central mode frequencies, $\omega_{0}^{(a)}$ and $\omega_{0}^{(b)}$. We couple the fluorescence to each filter through a 50:50 beam splitter, with one arm coupling to the first detection filter, \textit{A}, and the other arm into the second filter, \textit{B}. We then have the Hamiltonian
\begin{align}
  H &= \hbar \frac{ \Omega }{ 2 } \left( \sigma_{+} + \sigma_{-} \right) + \hbar \sum_{j=-N}^{N} \Delta\omega_{j}^{(a)} a_{j}^{\dagger} a_{j} \nonumber \\
  &\quad + \frac{ i \hbar }{ 2 } \sum_{j=-N}^{N} \left( \mathcal{E}_{j}^{*} a_{j} \sigma_{+} - \mathcal{E}_{j} a_{j}^{\dagger} \sigma_{-} \right)  \nonumber \\
  &\quad + \hbar \sum_{j=-N}^{N} \Delta\omega_{j}^{(b)} b_{j}^{\dagger} b_{j} \nonumber \\
  &\quad + \frac{ i \hbar }{ 2 } \sum_{j=-N}^{N} \left( \mathcal{E}_{j}^{*} b_{j} \sigma_{+} - \mathcal{E}_{j} b_{j}^{\dagger} \sigma_{-} \right) ,
\end{align}
and master equation
\begin{align}
  \ddt[\rho] &= \frac{ 1 }{ i \hbar } [ H , \rho ] + \frac{ \kappa }{ 2 } \sum_{j=-N}^{N} \Lambda \left( a_{j} \right) \rho + \frac{ 1 }{ 2 } \sum_{j=-N}^{N} \Lambda \left( C_{j}^{(a)} \right) \rho \nonumber \\
  &\quad + \frac{ \kappa }{ 2 } \sum_{j=-N}^{N} \Lambda \left( b_{j} \right) \rho + \frac{ 1 }{ 2 } \sum_{j=-N}^{N} \Lambda \left( C_{j}^{(b)} \right) \rho .
\end{align}
where
\begin{subequations}
  \begin{gather}
    \Delta\omega_{j}^{(a)} = \left( \omega_{0}^{(a)} + j \delta\omega \right) , \\
    \Delta\omega_{j}^{(b)} = \left( \omega_{0}^{(b)} + j \delta\omega \right) , \\
    \mathcal{E}_{j} = \sqrt{ \frac{ \gamma \kappa / 2 }{ 2 N + 1 } } e^{i m j \pi / N} , \label{eq:5_Ej_two_filter} \\
    C_{j}^{(a)} = \sqrt{ \frac{ \gamma / 2 }{ 2 N + 1 } } e^{i m j \pi / N} \sigma_{-} + \sqrt{ \kappa } a_{j} , \\
    C_{j}^{(b)} = \sqrt{ \frac{ \gamma / 2 }{ 2 N + 1 } } e^{i m j \pi / N} \sigma_{-} + \sqrt{ \kappa } b_{j} .
  \end{gather}
\end{subequations}
We note here the difference between Eqs.~(\ref{eq:5_Ej_two_filter}) and (\ref{eq:4_Ej_one_filter}) is the factor of $1/2$ to account for the 50:50 beam splitter guiding the fluorescence into the \textit{two} multi-mode array filters. For this two-filter system, we wish to correlate photons emitted from filter \textit{B} some time $\tau$ after a detection of an emission from mode \textit{A}. We therefore introduce the \textit{frequency-filtered second-order cross-correlation function}:
\begin{figure*}[t]
  \centering
  \includegraphics[width=\linewidth]{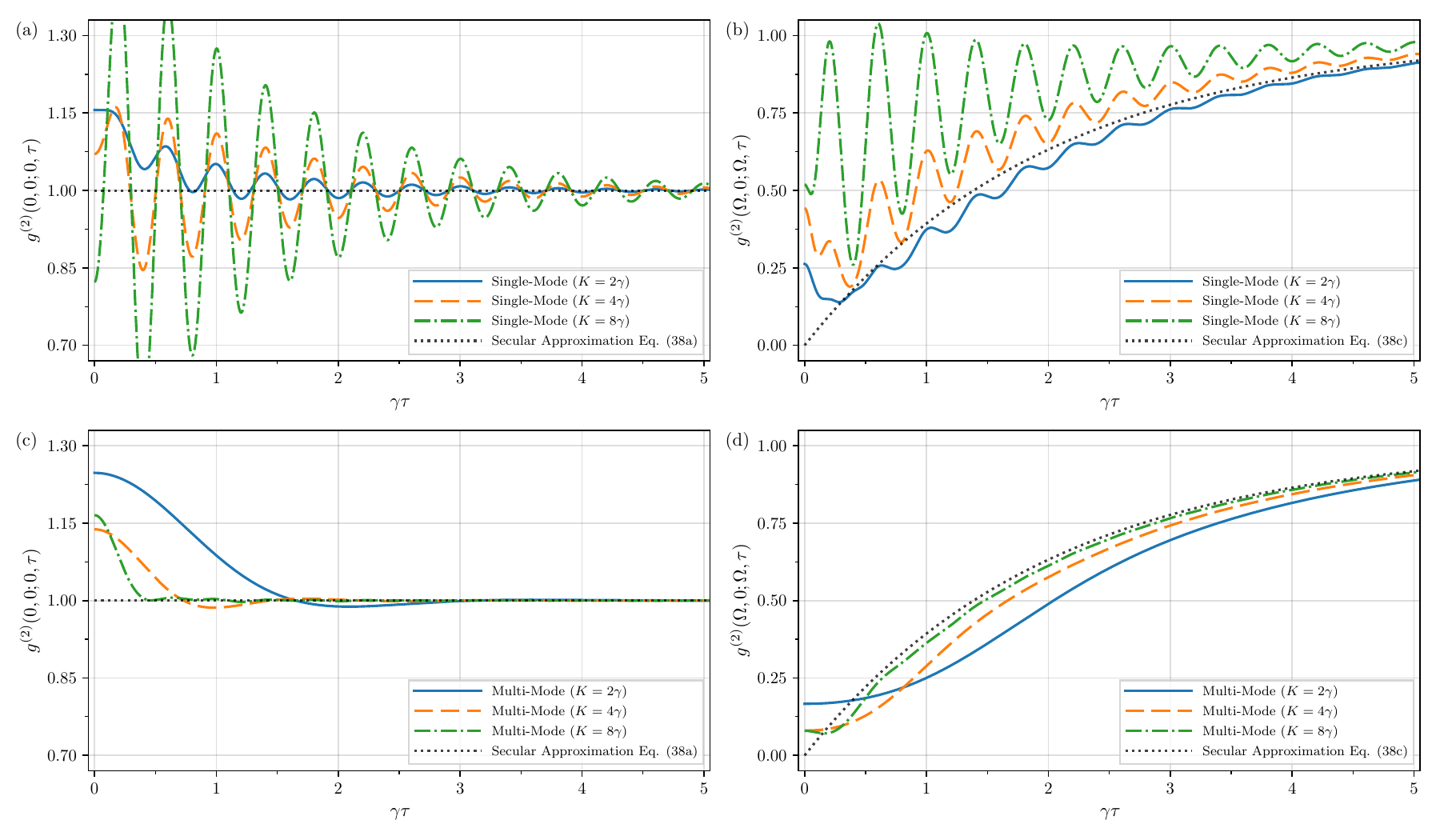}
  \caption{Frequency-filtered auto-correlation function of a single-mode (a, b), $N = 0$, and multi-mode (c, d), $N = 80$, filter. The filter resonance is resonant with the (a, c) central-peak, $\Delta\omega_{0} = 0$,  and (b, d) right-peak, $\Delta\omega_{0} = 5 \pi \gamma$, of the Mollow triplet, with halfwidths $K = 2 \gamma$ (blue, solid), $K = 4 \gamma$ (orange, dashed), and $K = 8 \gamma$ (green, dot-dash). The filtered correlation functions are plotted against the secular approximation equations, (a, b) Eq.~(\ref{eq:3_g2_dressed_C}) and (c, d) Eq.~(\ref{eq:3_g2_dressed_R}). Other parameters are $\Omega = 5 \pi \gamma, m = 1$ and, for the multi-mode array filter $\delta\omega = K / N, \kappa = 2.5 \delta\omega$.}
  \label{fig:5_combined_g2}
\end{figure*}
\begin{equation}\label{eq:5_g2_cross_equation}
  g^{(2)}(\alpha, 0; \beta, \tau) = \frac{ \langle A^{\dagger}(0) B^{\dagger} B(\tau) A(0) \rangle_{ss} }{ \langle A^{\dagger} A \rangle_{ss} \langle B^{\dagger} B \rangle_{ss} }
\end{equation}
where $\alpha$ and $\beta$ indicate the central resonance frequencies of filters $A$ and $B$, with respective collective annihilation operators,
\begin{equation}
  A = \sum_{j=-N}^{N} a_{j}, \quad B = \sum_{j=-N}^{N} b_{j} .
\end{equation}

We again use the moment equation approach as mentioned in the previous section, with the moment equations given the Appendix.

%------------------------------------------------------------------------------%
%-------------------------------%
%     5.1 Auto-correlations     %
%-------------------------------%
\subsection{Auto-correlations}
%------------------------------------------------------------------------------%

Figure~\ref{fig:4_filtered_spectrum} shows exactly what we expect of the multi-mode array filter in the frequency domain, that is, the multi-mode array filter is able isolate frequency components much more effectively than the single-mode filter. We now move into the time domain and towards the main focus of this paper: \textit{frequency-filtered photon correlations}. We therefore consider the case where $\alpha = \beta$ in Eq.~(\ref{eq:5_g2_cross_equation}).

There are four parameters of the multi-mode array filter: the number of modes, $N$; the frequency spacing between each mode, $\delta\omega$; the halfwidth of each individual mode, $\kappa$; and, the size of the phase modulation, $m$. For this work, however, we will fix the number of modes to $N = 80$, and set the frequency spacing based on the effective halfwidth, $K = N \delta\omega$. As established in Section~\ref{sect:2_filtering}, we will also assume the halfwidth of each mode to be much smaller than the effective halfwidth, with $\kappa = \frac{ \delta\omega }{ 4 } = \frac{ K }{ 4 N }$. We will also set the size of the phase modulation to $m = 1$. For this paper we will only consider the effect that the effective halfwidth, $K$, has on the single- and multi-mode-filtered correlation functions. For a detailed analysis on the effect of other parameters, see Ref.~\cite{Ngaha_2023_Frequency_Filtered_Photon_Correlations}, Sec.~6.3.1.

%------------------------------------------------------------------------------%
%-------------------------------------%
%     5.1.1 Varying the halfwidth     %
%-------------------------------------%
\subsubsection{Varying the halfwidth}
%------------------------------------------------------------------------------%

In Fig.~\ref{fig:5_combined_g2} we depict sets of frequency-filtered photon correlation functions for increasing halfwidths -- $K = 2 \gamma$ (blue solid line), $4 \gamma$ (orange dashed line), and $8 \gamma$ (green dot-dash line) -- for both the single-mode filter, Figs.~\ref{fig:5_combined_g2}(a) and \ref{fig:5_combined_g2}(b), and the multi-mode array filter, Figs.~\ref{fig:5_combined_g2}(c) \ref{fig:5_combined_g2}(d). We compare the frequency-filtered second-order correlation functions to the ideal correlation functions derived in the secular approximation (grey dotted line), Eqs.~(\ref{eq:3_g2_dressed_C}) and (\ref{eq:3_g2_dressed_R}).

\begin{figure}
  \centering
  \includegraphics[width=\linewidth]{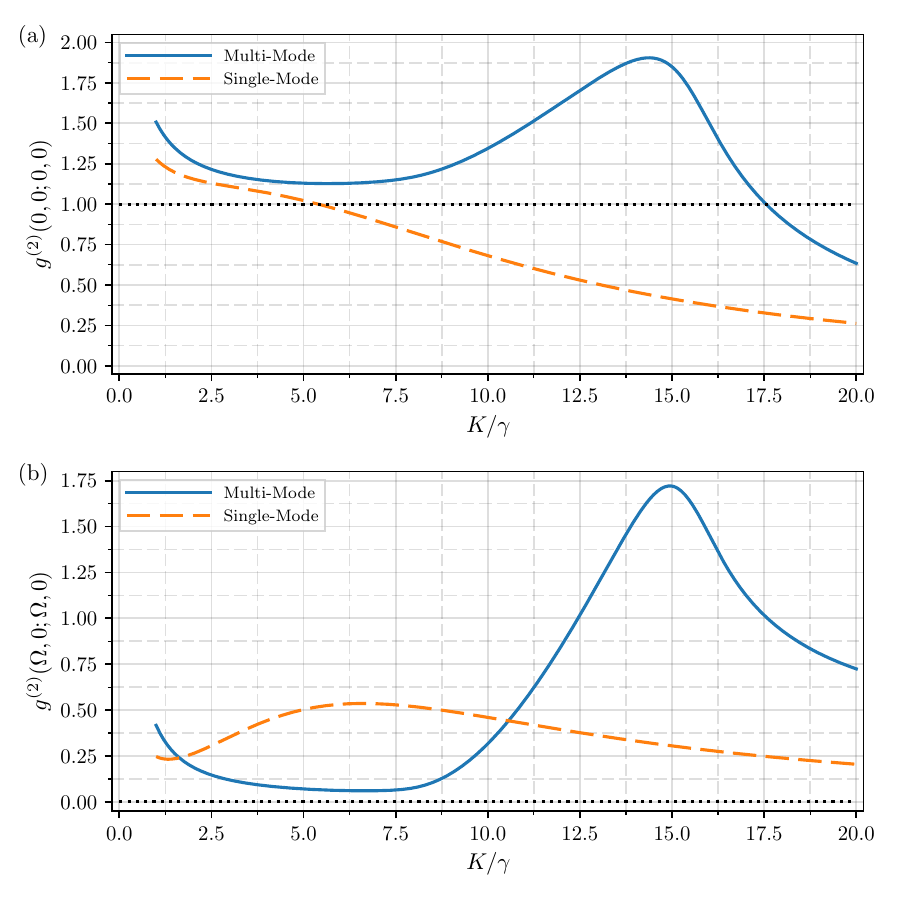}
  \caption{Initial photon-correlation value of the multi-mode (blue, solid), $N = 80$, and single-mode (orange, dashed), $N = 0$, filters for the (a) central peak, $\Delta\omega_{0} = 0$, and (b) right peak, $\Delta\omega_{0} = 5 \pi \gamma$, of the Mollow triplet. Also shown is the initial auto-correlation value from the secular approximations (black, dotted line), (a) $g^{(2)}(0, 0; 0, 0) = 1$ and (b) $g^{(2)}(\Omega, 0; \Omega, 0)$. Other parameters are $\Omega = 5 \pi \gamma, \delta\omega = K / N$, and $\kappa = 2.5 \delta\omega$.}
  \label{fig:5_g2_width_initial}
\end{figure}

In the bad cavity limit, as $K / \gamma \rightarrow \infty$, the ``filtered'' field is indistinguishable from the unfiltered atomic fluorescence; the photon correlations exhibit large Rabi oscillations, of frequency $\Omega$, as the atomic state oscillates between the ground and excited states. For the single-mode filter, Figs.~\ref{fig:5_combined_g2}(a) and \ref{fig:5_combined_g2}(b), these Rabi oscillations are visible even at the smaller halfwidth, $K = 2 \gamma$. As the halfwidth is increased, these oscillations get larger; indicating, in the temporal domain, that the frequency isolation is worsening.

We see an almost immediate improvement in the multi-mode array filter where, even for the largest halfwidth of $K = 8 \gamma$, the Rabi oscillations are almost completely suppressed. For $K = 8 \gamma$ we also see a much closer agreement with the secular approximations, Eqs.~(\ref{eq:3_g2_dressed_C}) and (\ref{eq:3_g2_dressed_R}). For smaller bandwidths, when $K \sim \gamma$, there is a larger discrepancy between the filtered correlations and the secular approximation, due to the slower temporal resolution of the finer bandwidth.

One gauge we can use to measure the efficacy of the multi-mode array filter is to investigate the initial value of the auto-correlation function. For very large filter bandwidths we expect an initial value close to zero, and for smaller bandwidths we expect to see different behaviour from the secular approximation. In Fig.~\ref{fig:5_g2_width_initial} we see the initial auto-correlation value tends towards zero as the filter halfwidth increases, while for smaller halfwidths it increases.

There is an intermediate region, $K \sim 2.5\gamma$ to $\sim 8\gamma$, where the initial correlation value approaches the value expected from the secular approximation: $g^{(2)}(0, 0; 0, 0) = 1$ for the central peak, Fig.~\ref{fig:5_g2_width_initial}(a); and $g^{(2)}(\Omega, 0; \Omega, 0) = 0$ for the right peak, Fig.~\ref{fig:5_g2_width_initial}(b). At first glance, in Fig.~\ref{fig:5_g2_width_initial}, it seems that the single-mode filter would perform better than the multi-mode array filter, as there is a direct match with the initial auto-correlation value in the secular approximation at $K \sim 5.5\gamma$. We have seen, however, in Figs.~\ref{fig:4_filtered_spectrum} and \ref{fig:5_combined_g2} that this halfwidth is still too large to effectively isolate the central peak.

There is a different story for the filtered right peak, Fig.~\ref{fig:5_combined_g2}(b). Heisenberg's uncertainty principal does not allow for a finite bandwidth filter to perfectly isolate a target photon with exact precision of its temporal correlations. In this halfwidth region, however, the multi-mode array filter produces stronger antibunching of the right peak when compared with the single-mode filter.

\begin{figure}
  \centering
  \includegraphics[width=\linewidth]{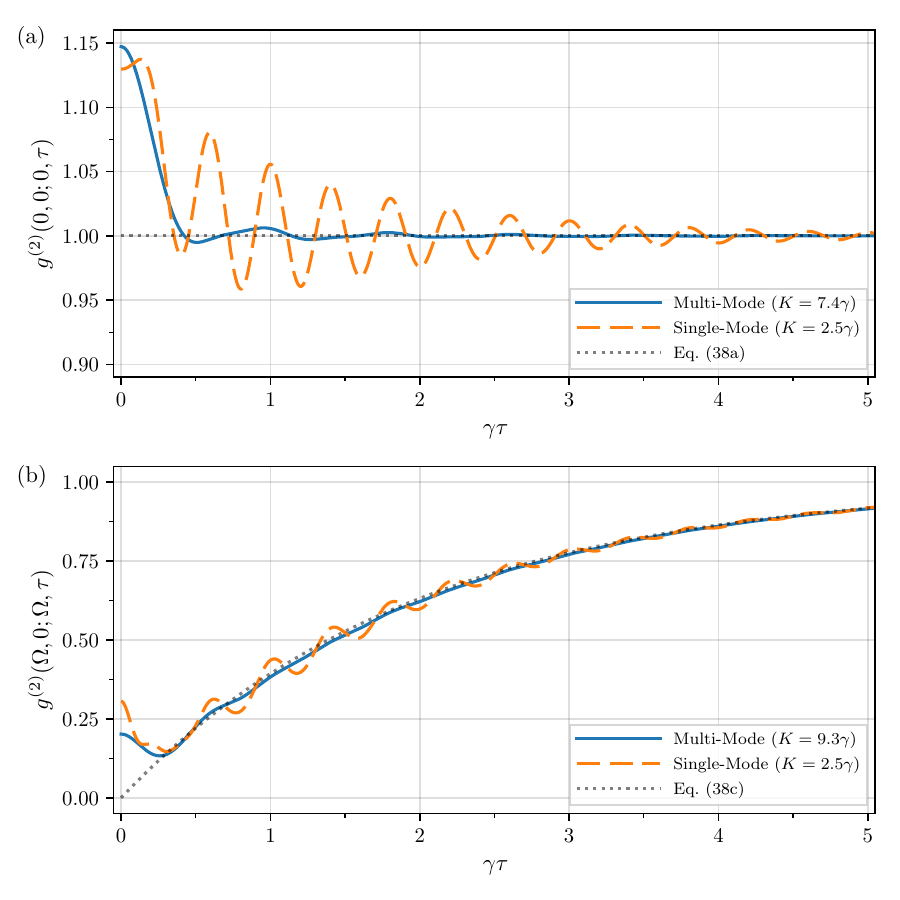}
  \caption{Frequency-filtered photon correlations of the (a) central peak, $\Delta\omega_{0} = 0$ and the (b) right peak, $\Delta\omega_{0} = 5 \pi \gamma$, of the Mollow triplet for the multi-mode (blue, solid), with $N = 80$, and single-mode (orange, dashed), with $N = 0$, filters. The filtered correlation functions are plotted against the secular approximation equations, (a) Eq.~(\ref{eq:3_g2_dressed_C}) and (b) Eq.~(\ref{eq:3_g2_dressed_R}). Other parameters are $\Omega = 5 \pi \gamma, \delta\omega = K / N$, and $\kappa = 2.5 \delta\omega$.}
  \label{fig:5_g2_good_vs_bad}
\end{figure}

In Fig.~\ref{fig:5_g2_good_vs_bad} we present auto-correlations calculated with single-mode and multi-mode array halfwidths that closely resemble the secular approximation correlation functions: $K = 7.4 \gamma~(2.5 \gamma)$ and $K = 9.3 \gamma~(2.5 \gamma)$ for the multi-mode~(single-mode) filtered central and right peak, respectively. Here we see that the multi-mode array filter has almost completely isolated the target peak. With an effective halfwidth roughly three times larger than the single-mode filter the Rabi oscillations are significantly suppressed, resulting in close matching of photon correlations to the secular approximations.

We note here that the filter halfwidths in Fig.~\ref{fig:5_g2_good_vs_bad} were chosen via numerical optimisation where the Rabi oscillations were minimised, i.e, a close match to the secular approximations. In principle, one could also just pick a halfwidth for the multi-mode array filter that is half the frequency separation between the target peak and its nearest neighbour, e.g., $\Omega / 2$ for the resonantly driven two-level atom, without substantially affecting the results.

%------------------------------------------------------------------------------%
%-----------------------------------------%
%     5.1.2 Filtering within the peak     %
%-----------------------------------------%
\subsubsection{Filtering within the peak}
%------------------------------------------------------------------------------%

\begin{figure}
  \centering
  \includegraphics[width=\linewidth]{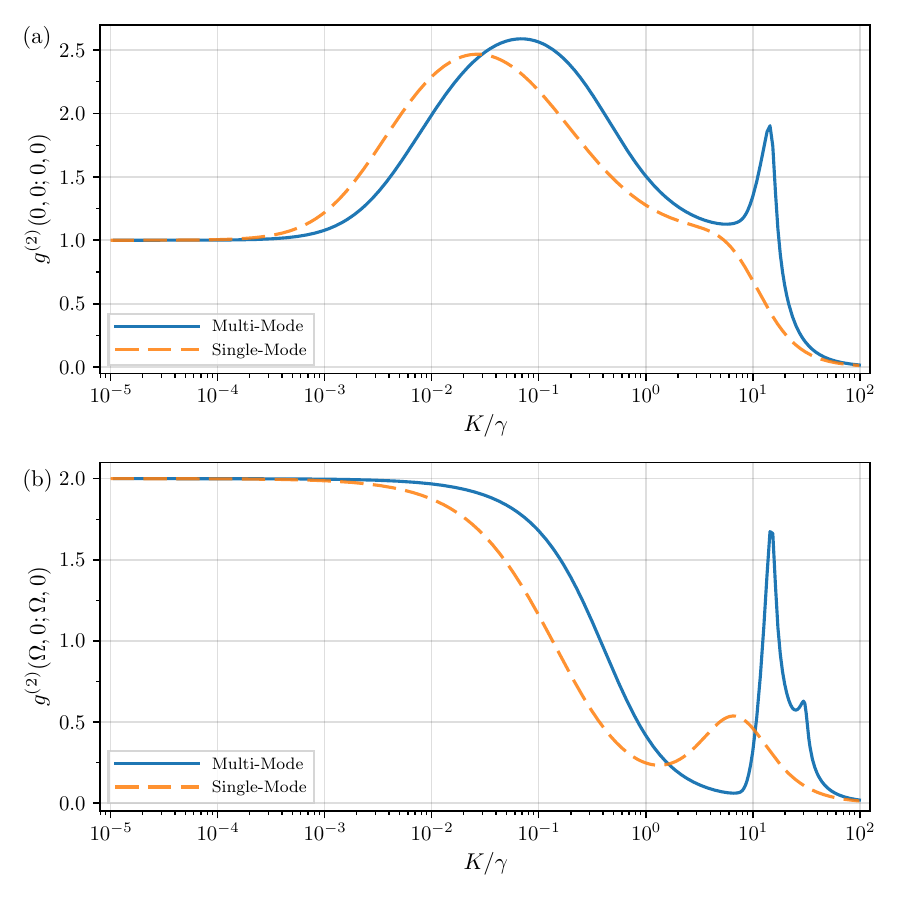}
  \caption{Initial value of the correlation function of the (a) central peak, $\Delta\omega_{0} = 0$ and (b) right peak, $\Delta\omega_{0} = 5 \pi \gamma$, for the multi-mode (blue, solid), $N = 80$, and single-mode (orange, dashed), $N = 0$, filters. Other parameters are $\Omega = 5 \pi \gamma, \delta\omega = K / N$, and $\kappa = 2.5 \delta\omega$.}
  \label{fig:5_g2_initial_scan}
\end{figure}

Figure \ref{fig:5_g2_width_initial} showed an increasing filter halfwidth resulted in auto-correlations that tended towards the bare atomic correlations. A decreasing halfwidth, however, resulted in correlations that deviated from the expected behaviour of the secular approximation. In Fig.~\ref{fig:5_g2_initial_scan} we extend the range of filter halfwidths from $K = 10^{-5} \gamma$ to $10^{2} \gamma$, highlighting the different regimes of behaviour. Results for the single-mode filter (orange dashed line) have been reported in both theoretical and experimental work in the strong \cite{Carreno_2022_PRA_LossAntibunching, GonzalezTudela_2015_PRA_OptimizationPhotonCorrelations} and weak driving regimes \cite{Phillips_2020_PRL_PhotonStatisticsFiltered, Phillips_2021__PhotonStatisticsFiltered, Hanschke_2020_PRL_OriginAntibunchingResonance}.

Focusing first on the single-mode filtered-correlations, we see similar regimes for both the filtered central and right-peaks [Figs.~\ref{fig:5_g2_initial_scan}(a) and \ref{fig:5_g2_initial_scan}(b), respectively]. As previously mentioned, very large filter halfwidths, $K \gg \Omega$, have a negligible effect on the atomic fluorescence, resulting in near perfect antibunching. As the halfwidth decreases to the range $\gamma < K < \Omega$, the filter mostly rejects the non-target peaks, resulting in auto-correlations close to the secular approximation, as discussed in the previous section.

As the halfwidth decreases further, we see two different, yet related, regimes for the filtered central- and right-peaks. For vanishingly small halfwidths, $K \lll \gamma$ ($\sim 10^{-5} \gamma$), the input field is effectively a broadband field from the perspective of the filter. The input field then begins to resemble a thermal environment, resulting in the ``thermalisation'' of the source field \cite{Nienhuis_1993_PRA_SpectralCorrelationsResonance, Joosten_2000_JoOBQaSO_InfluenceSpectralFiltering, GonzalezTudela_2013_NJoP_TwoPhotonSpectra}. The side-peaks of the Mollow triplet are products of the incoherent scattering of the driving field by the atom. Hence as the filter halfwidth decreases, the initial value of the auto-correlation function tends towards that of thermal light, $g^{(2)}(\Omega, 0; \Omega, 0) = 2$.

The central-peak differs, however, as there exists a coherent scattering component, which can be reasonably modelled by the Dirac-delta function. For vanishingly small halfwidths, more of the incoherent scattering is filtered out, eventually leaving only the infinitely narrow coherent component. The auto-correlation then tends towards second-order coherence, with $g^{(2)}(0, 0; 0, \tau) = 1$.

The two possible de-excitation paths corresponding to the central peak (see Fig.~\ref{fig:3_2LA_dressed_states}(a)) should give rise to photon bunching, however, interference between these two transitions results in second-order coherence \cite{GonzalezTudela_2013_NJoP_TwoPhotonSpectra}. For small -- yet not vanishingly small -- halfwidths, $K \sim 10^{-2}\gamma - 10^{-1} \gamma$, this interference no longer holds and we see a large peak in photon bunching; $g^{(2)}(0, 0; 0, 0) \approx 2.5$ for the single-mode filter.

The multi-mode array filter produces results similar to the single-mode filter: large halfwidths give near perfect antibunching, intermediate halfwidths can isolate the target peaks, and vanishingly small halfwidths result in thermalisation for the filtered right peak and second-order coherence for the filtered central peak. We do see some slight differences, such as a shift in the bunching peak for the filtered central peak, due to the improved frequency isolation. The most striking difference is the emergence of peaks at $K \sim \Omega$ for both the filtered central- and right-peaks, and a second peak at $K \sim 2 \Omega$ for the filtered right-peak. Similar results can be seen in a perfect rectangular filter (see Figs.~4 and 5 of Ref.~\cite{Kamide_2015_PRA_EigenvalueDecompositionMethod}). Kamide et al. suggest that these peaks are due to constructive interference in the cascaded decay channels.

\begin{figure}
  \centering
  \includegraphics[width=\linewidth]{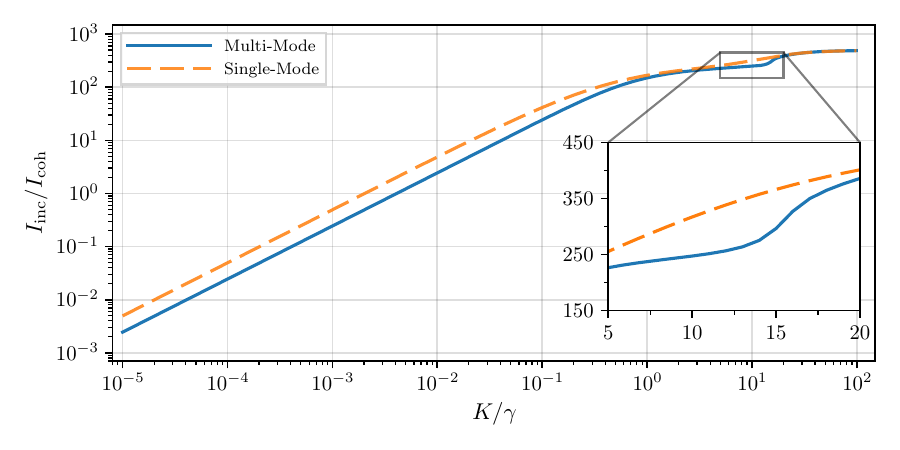}
  \caption{Intensity ratio of the multi-mode (blue, solid) and single-mode (orange, dashed) filtered coherent and incoherent spectra when resonant with the central peak, $\Delta\omega = 0$, from Eq.~(\ref{eq:5_intensity_ratio}). The inset is zoomed in on the region where $K \approx \Omega$, where there is a clear decrease in the incoherent intensity $I_{\mathrm{inc}}$, corresponding to the bunching peak in Fig.~\ref{fig:5_g2_initial_scan}. Parameters are the same as in Fig.~\ref{fig:5_g2_width_initial}.}
  \label{fig:5_intensity_ratios}
\end{figure}

The ratio of the coherent and incoherent scattering intensities for the filtered field is given by
\begin{equation}\label{eq:5_intensity_ratio}
  \frac{ I_{\mathrm{inc}} }{ I_{\mathrm{coh}} } = \frac{ \langle \Delta A^{\dagger} \Delta A \rangle_{ss} }{ \langle A^{\dagger} \rangle_{ss} \langle A \rangle_{ss} } = \frac{ \langle A^{\dagger} A \rangle_{ss} }{ \langle A^{\dagger} \rangle_{ss} \langle A \rangle_{ss} } - 1 ,
\end{equation}
which we show for the filtered central peak in Fig.~\ref{fig:5_intensity_ratios}. When the halfwidth is approximately the same as the side-peak splitting, $K \approx \Omega$, there is a clear deviation in the single-mode and multi-mode filtered intensity ratios (highlighted in the inset of Fig.~\ref{fig:5_intensity_ratios}). As the frequency response of the multi-mode array filter expands and overlaps slightly with the sidepeaks, there is a decrease in the incoherent scattering intensity, possibly due to the interference in the cascaded decay channels. We can infer a similar effect occurs when the filter is resonant with either sidepeak. As the filter response first overlaps with the central peak, at $K \approx \Omega$, and then with the opposite side peak, at $K \approx 2 \Omega$, there will be a similar decrease in the incoherent scattering. This, however, will need further investigation.

%------------------------------------------------------------------------------%
%--------------------------------%
%     5.2 Cross-correlations     %
%--------------------------------%
\subsection{Cross-correlations}
%------------------------------------------------------------------------------%

Allowing the tunable central frequencies of both filters, $\Delta\omega_{0}^{(a)}$ and $\Delta\omega_{0}^{(b)}$, to vary, we can investigate \textit{cross-correlations} between different frequency components.

\begin{figure}
  \centering
  \includegraphics[width=\linewidth]{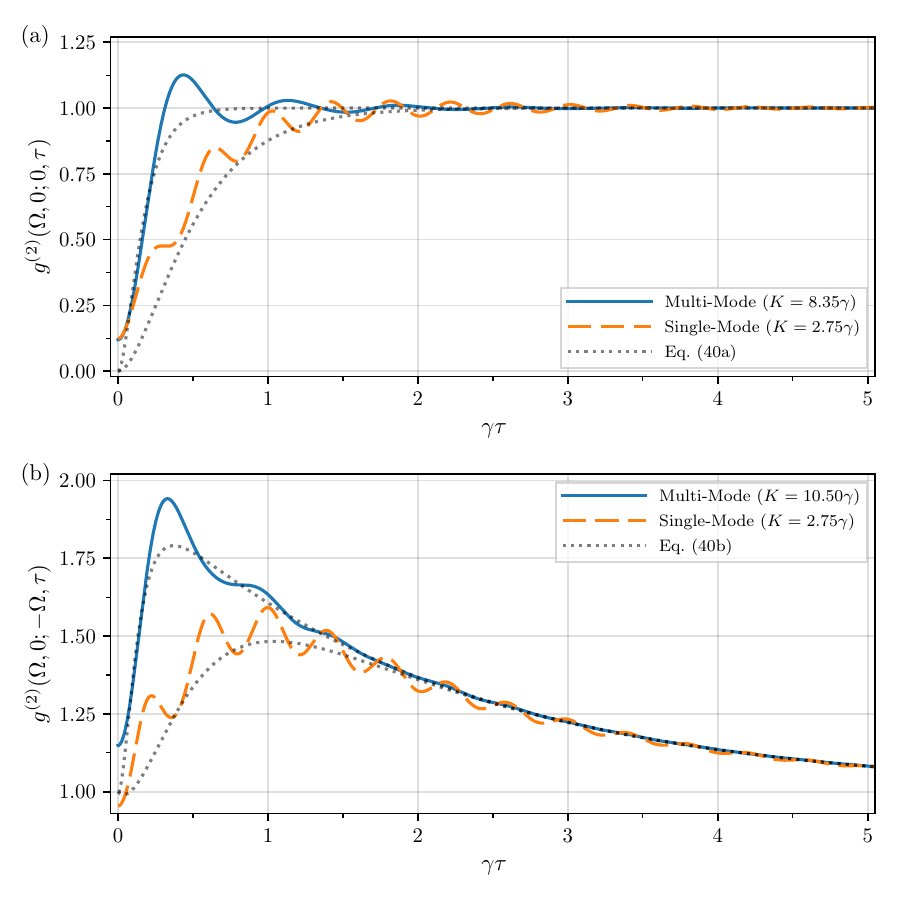}
  \caption{Cross-correlation function of the filtered (a) right-to-central peak, $g^{(2)}(\Omega, 0; 0, \tau)$, and (b) right-to-left-peak, $g^{(2)}(\Omega, 0; -\Omega, \tau)$, transitions for the multi-mode (blue, solid) and single-mode (orange, dashed) filters. The filtered correlation functions are plotted against the secular approximation, Eq.~(\ref{eq:3_g2_dressed_RL}), for the respective halfwidth. Other parameters are $\Omega = 5 \pi \gamma, \delta\omega = K / N$, and $\kappa = 2.5 \delta\omega$.}
  \label{fig:5_combined_g2_cross}
\end{figure}

\begin{figure*}[t]
  \centering
  \includegraphics[width=\linewidth]{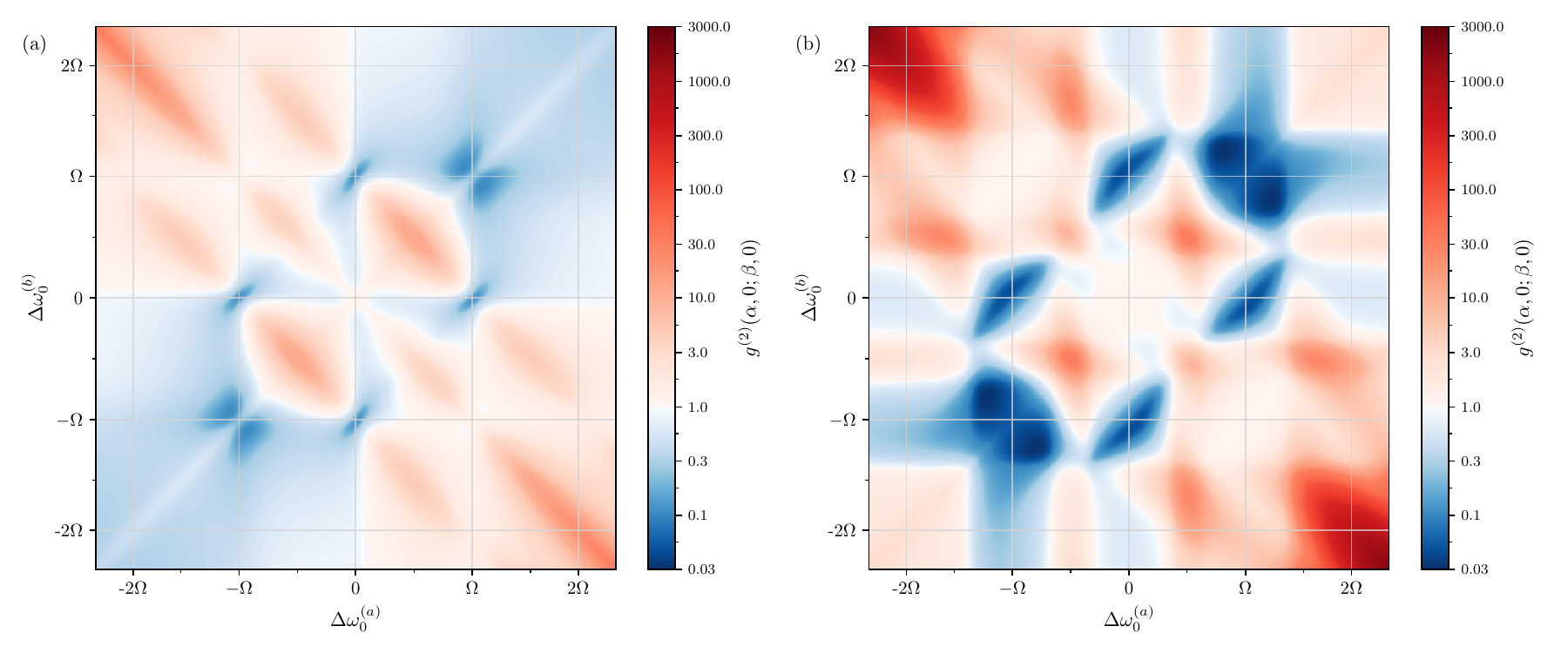}
  \caption{Initial value of the cross-correlation function, $g^{(2)}(\alpha, 0; \beta, 0)$, for the (a) single-mode filter, with $N = 0$ and $K = \gamma$, and the (b) multi-mode array filters, with $N = 80, K = 5.5 \gamma, \delta\omega = 0.069 \gamma$, and $\kappa = 0.172 \gamma$. The colour code of the contour plot is: blue for $g^{(2)}(\alpha, 0; \beta, 0) < 1$, white for $g^{(2)}(\alpha, 0; \beta, 0) = 1$, and red for $g^{(2)}(\alpha, 0; \beta, 0) > 1$. The driving amplitude is $\Omega = 5 \pi \gamma$}
  \label{fig:5_combined_cross_scan}
\end{figure*}

\subsubsection{Temporal evolution of peak-to-peak correlations}

We first focus on the temporal evolution of the cross-correlation function when the two filters are resonant with two different peaks of the Mollow triplet. When driven on resonance, there is a symmetry between the two side-peaks, therefore we set the first filter, $A$, resonant with the right-peak, $\Delta\omega_{0}^{(a)} = \Omega$. In Fig.~\ref{fig:5_combined_g2_cross} we depict the evolution of the (a) right-to-central peak, $g^{(2)}(\Omega, 0; 0, \tau)$, and (b) right-to-left peak, $g^{(2)}(\Omega, 0; -\Omega, \tau)$, cross-correlations.

As previously discussed, the secular approximation in the atomic picture fails to account for the different time orderings of the de-excitation paths. The filtered correlations in Fig.~\ref{fig:5_combined_g2_cross} do not show the expected bunching for right-to-left filtered-correlations of Eq.~(\ref{eq:3_dressed_state_cross_corr_RC}), or the second-order coherence of Eq.~(\ref{eq:3_dressed_state_cross_corr_RL}). We instead see close agreement with the behaviour expected from the equations derived by Schrama et al. \cite{Schrama_1991_PRL_DestructiveInterferenceOpposite}, Eqs.~(\ref{eq:3_schrama_cross_eqns}).

These theoretical results do slightly differ from experimental literature, as in Refs.~\cite{Aspect_1980_PRL_TimeCorrelationsTwo, Schrama_1992_PRA_IntensityCorrelationsComponents}. For the right-to-central peak cross-correlation, Fig.~\ref{fig:5_combined_g2_cross}a, the central dip due to the destructive interference is seen in Figs.~12-14 of Ref.~\cite{Schrama_1992_PRA_IntensityCorrelationsComponents}. The experimental results, however, do not show any visible Rabi oscillations, possibly due to the weaker driving amplitude.

We note here that we have chosen different halfwidths for the single- and multi-mode array filters, with the same numerical optimisation method as in Fig.~\ref{fig:5_g2_good_vs_bad}, which is why the expected short-time behaviours differ. With a halfwidth roughly three times larger, however, the visible Rabi oscillations in the multi-mode-filtered cross-correlations are still significantly reduced when compared to the single-mode filter. Larger $K$ for the single-mode filter would result in much larger Rabi oscillations in the cross-correlations.

\subsubsection{Cross-correlations outside the Mollow triplet peaks}

We are not limited to cross-correlations that are resonant with the three peaks of the Mollow triplet; we can choose the central resonance of filters $A$ and $B$ to be any frequency. Doing so allows us to uncover more interesting, and potentially useful, photon correlations. Figure~\ref{fig:5_combined_cross_scan} depicts landscapes of initial correlation values for the single- and multi-mode array filters, as the central frequencies of both filters are varied. We note that the results for the single-mode filter, Fig.~\ref{fig:5_combined_cross_scan}(a) closely resembles the results from recent work \cite{GonzalezTudela_2013_NJoP_TwoPhotonSpectra, GonzalezTudela_2015_PRA_OptimizationPhotonCorrelations, Carreno_2017_L&PR_PhotonCorrelationsMollow, Munoz_2014_PRA_ViolationClassicalInequalities, Laussy_2017_NM_NewWayCorrelate}, where the frequency filter was also modelled as a single-mode cavity.

For both filtering cases we see similar regions of correlations. Along the positive diagonal -- the initial autocorrelation value -- we see regions of antibunching surrounding the side-peaks and second-order coherence about the origin. For cross-correlated photons, there are regions of strong antibunching when one of the filters is resonant with the central peak, and second-order coherence for the left-to-right and right-to-left peak correlations. These regions are anticipated from the temporal evolution, Figs.~\ref{fig:5_combined_g2} and \ref{fig:5_combined_g2_cross}, and from the secular approximations, Eqs.~(\ref{eq:3_dressed_state_auto_correlation_eqns}) and (\ref{eq:3_schrama_cross_eqns}).

When moving the filter resonances away from the Mollow triplet peaks, we see regions of strong photon correlations appearing. These regions are due to simultaneous two-photon ``leapfrog'' processes. As there are only three possible relaxation paths, these regions occur in areas where the sum of the two filtered photons' frequencies matches the allowed transition, i.e., $\Delta\omega_{0}^{(a)} + \Delta\omega_{0}^{(b)} = -\Omega, 0$, or $\Omega$. Due to the improved frequency isolation of the multi-mode array filter, these regions are much more clearly defined.

%==============================================================================%
%                                 6 CONCLUSION                                 %
%==============================================================================%
\section{Conclusion}
\label{sect:6_conclusion}

We have developed an effective frequency filtering model -- the \textit{multi-mode array filter} -- consisting of an array of tunable single-mode cavities. Each cavity is assumed to have a very narrow linewidth compared to the range of frequencies, such that there is minimal overlap between each individual frequency response. The output of each individual cavity is then directed towards a single photodetector, resulting in a near-rectangular collective frequency response. Compared against a single-mode Lorentzian filter, the multi-mode array filter can achieve much wider halfwidths -- and, thus, faster temporal response -- with a much sharper cut-off in frequency response.

We investigated frequency-filtered photon correlations when filtering resonance fluorescence of a driven two-level atom. We compared results for frequency-filtered auto- and cross-correlations for both the single-mode and multi-mode filters, where the multi-mode array-filtered correlations could reproduce the ideal photon correlations, as derived in the secular approximation, more accurately than the single-mode filter.

In order to achieve a sufficient frequency response, a large number of cavity modes, $N$, are required. This would subsequently result in a rapidly growing Hilbert space of the model. Due to the cascaded coupling, we were able to circumvent this issue by deriving a closed set of coupled operator moment equations, providing an efficient method of calculation. Provided one can derive a closed set of operator moment equations for the source system, cascading the source into the multi-mode array ensures a closed set of equations for the total source-filter coupled system. Unfortunately, not every quantum system can be described by a closed set of operator equations, in which case a master equation or quantum trajectory method must be used, with a conservative number of modes.

Here we have defined the multi-mode array filter as an array of single-mode resonators. While not the focus of this paper, recent work with ring resonator arrays \cite{Mirza_2013_JotOSoAB_SinglePhotonTime} and on-chip resonators \cite{Maring_2024_NP_VersatileSinglePhoton, Yard_2024_PRL_ChipQuantumInformation} show promise for possible experimental implementations of the multi-mode array filter.

In principle, this filtering system could be applied to any quantum optical source system. To calculate the complete set of filter-operator moment equations, the set of moment equations of the source system must be closed. We have also applied this method to a three-level ladder-type atom driven at two-photon resonance, with a manuscript currently in preparation \cite{Ngaha_3LA_unpublished}.

%==============================================================================%
%                               ACKNOWLEDGEMENTS                               %
%==============================================================================%

\begin{acknowledgments}
  The authors wish to acknowledge the use of New Zealand eScience Infrastructure (NeSI) high performance computing facilities, consulting support and/or training services as part of this research. New Zealand's national facilities are provided by NeSI and funded jointly by NeSI's collaborator institutions and through the Ministry of Business, Innovation \& Employment's Research Infrastructure programme.
\end{acknowledgments}

%==============================================================================%
%                                  APPENDICES                                  %
%==============================================================================%
\appendix

% Begin wide text for all appendix equations
\begin{widetext}
  \section{Moment Equations for the Two-Filter System}
  \label{app:moment_equations}

  Here we present all of the operator moment equations needed to calculate the second-order cross-correlation function, Eq.~(\ref{eq:5_g2_cross_equation}). We write these coupled moment equations in matrix form, where we use the notation:
  \begin{equation}
    \langle \bm{\sigma} \rangle =
    \begin{pmatrix}
      \langle \sigma_{-} \rangle \\
      \langle \sigma_{+} \rangle \\
      \langle \sigma_{z} \rangle
    \end{pmatrix}
    , \quad \langle X \bm{\sigma} \rangle =
    \begin{pmatrix}
      \langle X \sigma_{-} \rangle \\
      \langle X \sigma_{+} \rangle \\
      \langle X \sigma_{z} \rangle
    \end{pmatrix}
    .
  \end{equation}

  Routines written in \verb!Fortan90! compute the steady states of these moment equations using the \verb!LAPACK! matrix inversion routines \cite{AndersonEtAl_LapackUsersGuide}. The routines can be found in a GitHub repository \cite{Jacob_Github}.

  %------------------------------------------------------------------------------%
  \subsection*{Two-level atom}
  %------------------------------------------------------------------------------%
  We restate the moment equations for the driven two-level atom, the optical Bloch equations, Eq.~(\ref{eq:3_optical_bloch_equations}):
  \begin{equation}
    \ddt \langle \bm{\sigma} \rangle = \bm{M}^{(\bm{\sigma})} \langle \bm{\sigma} \rangle +
    \begin{pmatrix}
      0 \\
      0 \\
      -\gamma
    \end{pmatrix}
    , \quad\quad \bm{M}^{(\bm{\sigma})} =
    \begin{pmatrix}
      - \frac{\gamma}{2} & 0 & i \frac{\Omega}{2} \\
      0 & -\frac{\gamma}{2} & -i \frac{\Omega}{2} \\
      i \Omega & -i \Omega & -\gamma
    \end{pmatrix} .
  \end{equation}

  %------------------------------------------------------------------------------%
  \subsection*{First-order filter mode operators}
  %------------------------------------------------------------------------------%
  The operator moment equations for the first-order filter mode operators, $\langle a_{j} \rangle, \langle a_{j}^{\dagger} \rangle, \langle b_{j} \rangle$, and $\langle b_{j}^{\dagger} \rangle$, are:
  \begin{subequations}
    \begin{align}
      \ddt \langle a_{j} \rangle &= -\left( \kappa + i \Delta\omega_{j}^{(a)} \right) \langle a_{j} \rangle - \mathcal{E}_{j} \langle \sigma_{-} \rangle , \\
      \ddt \langle a_{j}^{\dagger} \rangle &= -\left( \kappa - i \Delta\omega_{j}^{(a)} \right) \langle a_{j}^{\dagger} \rangle - \mathcal{E}_{j}^{*} \langle \sigma_{+} \rangle , \\
      %---------------------------
      \ddt \langle b_{j} \rangle &= -\left( \kappa + i \Delta\omega_{j}^{(b)} \right) \langle b_{j} \rangle - \mathcal{E}_{j} \langle \sigma_{-} \rangle , \\
      \ddt \langle b_{j}^{\dagger} \rangle &=  -\left( \kappa - i \Delta\omega_{j}^{(b)} \right) \langle b_{j}^{\dagger} \rangle - \mathcal{E}_{j}^{*} \langle \sigma_{+} \rangle .
    \end{align}
  \end{subequations}

  %------------------------------------------------------------------------------%
  \subsection*{First-order filter mode operators with spin operators}
  %------------------------------------------------------------------------------%
  The operator moment equations for the second-order operators, $\langle a_{j} \bm{\sigma} \rangle, \langle a_{j}^{\dagger} \bm{\sigma} \rangle, \langle b_{j} \bm{\sigma} \rangle$, and $\langle b_{j}^{\dagger} \bm{\sigma} \rangle$ are:
  \begin{subequations}
    \begin{align}
      \ddt \langle a_{j} \bm{\sigma} \rangle &= \left[ \bm{M}^{(\sigma)} - \left( \kappa + i \Delta\omega_{j}^{(a)} \right) \mathds{1} \right] \langle a_{j} \bm{\sigma} \rangle +
      \begin{pmatrix}
        0 \\
        -\frac{ 1 }{ 2 } \mathcal{E}_{j} \left( \langle \sigma_{z} \rangle + 1 \right) \\
        -\gamma \langle a_{j} \rangle + \mathcal{E}_{j} \langle \sigma_{-} \rangle
      \end{pmatrix}
      , \\
      %--------------------------------------------%
      \ddt \langle a_{j}^{\dagger} \bm{\sigma} \rangle &= \left[ \bm{M}^{(\sigma)} - \left( \kappa - i \Delta\omega_{j}^{(a)} \right) \mathds{1} \right] \langle a_{j} \bm{\sigma} \rangle +
      \begin{pmatrix}
        -\frac{ 1 }{ 2 } \mathcal{E}_{j}^{*} \left( \langle \sigma_{z} \rangle + 1 \right) \\
        0 \\
        -\gamma \langle a_{j}^{\dagger} \rangle + \mathcal{E}_{j}^{*} \langle \sigma_{+} \rangle
      \end{pmatrix}
      , \\
      %--------------------------------------------%
      \ddt \langle b_{j} \bm{\sigma} \rangle &= \left[ \bm{M}^{(\sigma)} - \left( \kappa + i \Delta\omega_{j}^{(b)} \right) \mathds{1} \right] \langle b_{j} \bm{\sigma} \rangle +
      \begin{pmatrix}
        0 \\
        -\frac{ 1 }{ 2 } \mathcal{E}_{j} \left( \langle \sigma_{z} \rangle + 1 \right) \\
        -\gamma \langle b_{j} \rangle + \mathcal{E}_{j} \langle \sigma_{-} \rangle
      \end{pmatrix}
      , \\
      %--------------------------------------------%
      \ddt \langle b_{j}^{\dagger} \bm{\sigma} \rangle &= \left[ \bm{M}^{(\sigma)} - \left( \kappa - i \Delta\omega_{j}^{(b)} \right) \mathds{1} \right] \langle b_{j} \bm{\sigma} \rangle +
      \begin{pmatrix}
        -\frac{ 1 }{ 2 } \mathcal{E}_{j}^{*} \left( \langle \sigma_{z} \rangle + 1 \right) \\
        0 \\
        -\gamma \langle a_{j}^{\dagger} \rangle + \mathcal{E}_{j}^{*} \langle \sigma_{+} \rangle
      \end{pmatrix}
      .
    \end{align}
  \end{subequations}

  %------------------------------------------------------------------------------%
  \subsection*{Second-order filter mode operators}
  %------------------------------------------------------------------------------%
  \vspace{-0.3cm}
  The operator moment equations for the second-order operators, $\langle a_{j} b_{k} \rangle, \langle a_{j}^{\dagger} b_{k}^{\dagger} \rangle, \langle a_{j}^{\dagger} a_{k} \rangle, \langle b_{j}^{\dagger} b_{k} \rangle, \langle a_{j}^{\dagger} b_{k} \rangle$, and $\langle b_{j}^{\dagger} a_{k} \rangle$ are:
  \vspace{-0.3cm}
  \begin{subequations}
    \begin{align}
      \ddt \langle a_{j} b_{k} \rangle &= -\left[ 2 \kappa + i \left( \Delta\omega_{j}^{(a)} + \Delta\omega_{k}^{(b)} \right) \right] \langle a_{j} b_{k} \rangle - \mathcal{E}_{j} \langle b_{k} \sigma_{-} \rangle - \mathcal{E}_{k} \langle a_{j} \sigma_{-} \rangle , \\
      %--------------------------------------------%
      \ddt \langle a^{\dagger}_{j} b^{\dagger}_{k} \rangle &= -\left[ 2 \kappa - i \left( \Delta\omega_{j}^{(a)} + \Delta\omega_{k}^{(b)} \right) \right] \langle a^{\dagger}_{j} b^{\dagger}_{k} \rangle - \mathcal{E}_{j}^{*} \langle b_{k}^{\dagger} \sigma_{+} \rangle - \mathcal{E}_{k}^{*} \langle a_{j}^{\dagger} \sigma_{+} \rangle , \\
      %--------------------------------------------%
      \ddt \langle a^{\dagger}_{j} a_{k} \rangle &= -\left[ 2 \kappa - i \left( \Delta\omega_{j}^{(a)} - \Delta\omega_{k}^{(a)} \right) \right] \langle a^{\dagger}_{j} a_{k} \rangle - \mathcal{E}_{j}^{*} \langle a_{k} \sigma_{+} \rangle - \mathcal{E}_{k} \langle a_{j}^{\dagger} \sigma_{-} \rangle , \\
      %--------------------------------------------%
      \ddt \langle b^{\dagger}_{j} b_{k} \rangle &= -\left[ 2 \kappa - i \left( \Delta\omega_{j}^{(b)} - \Delta\omega_{k}^{(b)} \right) \right] \langle b^{\dagger}_{j} b_{k} \rangle - \mathcal{E}_{j}^{*} \langle b_{k} \sigma_{+} \rangle - \mathcal{E}_{k} \langle b_{j}^{\dagger} \sigma_{-} \rangle , \\
      %--------------------------------------------%
      \ddt \langle a^{\dagger}_{j} b_{k} \rangle &= -\left[ 2 \kappa - i \left( \Delta\omega_{j}^{(a)} - \Delta\omega_{k}^{(b)} \right) \right] \langle a^{\dagger}_{j} b_{k} \rangle - \mathcal{E}_{j}^{*} \langle b_{k} \sigma_{+} \rangle - \mathcal{E}_{k} \langle a_{j}^{\dagger} \sigma_{-} \rangle , \\
      %--------------------------------------------%
      \ddt \langle b^{\dagger}_{j} a_{k} \rangle &= -\left[ 2 \kappa - i \left( \Delta\omega_{j}^{(b)} - \Delta\omega_{k}^{(a)} \right) \right] \langle b^{\dagger}_{j} a_{k} \rangle - \mathcal{E}_{j}^{*} \langle a_{k} \sigma_{+} \rangle - \mathcal{E}_{k} \langle b_{j}^{\dagger} \sigma_{-} \rangle .
    \end{align}
  \end{subequations}
  %\vspace{-1cm}

  %------------------------------------------------------------------------------%
  \subsection*{Second-order filter mode operators with spin operators}
  %------------------------------------------------------------------------------%
  \vspace{-0.3cm}
  The operator moment equations for the third-order operators, $\langle a_{j} b_{k} \rangle, \langle a_{j}^{\dagger} b_{k}^{\dagger} \bm{\sigma} \rangle, \langle a_{j}^{\dagger} a_{k} \bm{\sigma} \rangle, \langle b_{j}^{\dagger} b_{k} \bm{\sigma} \rangle, \langle a_{j}^{\dagger} b_{k} \bm{\sigma} \rangle$, and $\langle b_{j}^{\dagger} a_{k} \bm{\sigma} \rangle$ are:
  \vspace{-0.3cm}
  \begin{subequations}
    \begin{align}
      \ddt \langle a_{j} b_{k} \bm{\sigma} \rangle &= \left\{ \bm{M}^{(\sigma)} - \left[ 2 \kappa + i \left( \Delta\omega_{j}^{(a)} + \Delta\omega_{k}^{(b)} \right) \right] \mathds{1} \right\} \langle a_{j} b_{k} \bm{\sigma} \rangle \nonumber \\
      &\quad\quad\quad\quad\quad\quad +
      \begin{pmatrix}
        0 \\
        -\frac{ 1 }{ 2 } \mathcal{E}_{j} \left( \langle b_{k} \sigma_{z} \rangle + \langle b_{k} \rangle \right) - \frac{ 1 }{ 2 } \mathcal{E}_{k} \left( \langle a_{j} \sigma_{z} \rangle + \langle a_{j} \rangle \right) \\
        -\gamma \langle a_{j} b_{k} \rangle + \mathcal{E}_{j} \langle b_{k} \sigma_{-} \rangle + \mathcal{E}_{k} \langle a_{j} \sigma_{-} \rangle
      \end{pmatrix}
      , \\
      %--------------------------------------------%
      \ddt \langle a^{\dagger}_{j} b^{\dagger}_{k} \bm{\sigma} \rangle &= \left\{ \bm{M}^{(\sigma)} - \left[ 2 \kappa - i \left( \Delta\omega_{j}^{(a)} + \Delta\omega_{k}^{(b)} \right) \right] \mathds{1} \right\} \langle a^{\dagger}_{j} b^{\dagger}_{k} \bm{\sigma} \rangle \nonumber \\
      &\quad\quad\quad\quad\quad\quad +
      \begin{pmatrix}
        -\frac{ 1 }{ 2 } \mathcal{E}_{j}^{*} \left( \langle b_{k}^{\dagger} \sigma_{z} \rangle + \langle b_{k}^{\dagger} \rangle \right) - \frac{ 1 }{ 2 } \mathcal{E}_{k}^{*} \left( \langle a_{j}^{\dagger} \sigma_{z} \rangle + \langle a_{j}^{\dagger} \rangle \right) \\
        0 \\
        -\gamma \langle a_{j}^{\dagger} b_{k}^{\dagger} \rangle + \mathcal{E}_{j}^{*} \langle b_{k}^{\dagger} \sigma_{+} \rangle + \mathcal{E}_{k}^{*} \langle a_{j}^{\dagger} \sigma_{+} \rangle
      \end{pmatrix}
      , \\
      %--------------------------------------------%
      \ddt \langle a^{\dagger}_{j} a_{k} \bm{\sigma} \rangle &= \left\{ \bm{M}^{(\sigma)} - \left[ 2 \kappa - i \left( \Delta\omega_{j}^{(a)} - \Delta\omega_{k}^{(a)} \right) \right] \mathds{1} \right\} \langle a^{\dagger}_{j} a_{k} \bm{\sigma} \rangle \nonumber \\
      &\quad\quad\quad\quad\quad\quad +
      \begin{pmatrix}
        -\frac{ 1 }{ 2 } \mathcal{E}_{j}^{*} \left( \langle a_{k} \sigma_{z} \rangle + \langle a_{k} \rangle \right) \\
        -\frac{ 1 }{ 2 } \mathcal{E}_{k} \left( \langle a_{j}^{\dagger} \sigma_{z} \rangle + \langle a_{j}^{\dagger} \rangle \right) \\
        -\gamma \langle a_{j}^{\dagger} a_{k} \rangle + \mathcal{E}_{j}^{*} \langle a_{k} \sigma_{+} \rangle + \mathcal{E}_{k} \langle a_{j}^{\dagger} \sigma_{-} \rangle
      \end{pmatrix}
      , \\
      %--------------------------------------------%
      \ddt \langle b^{\dagger}_{j} b_{k} \bm{\sigma} \rangle &= \left\{ \bm{M}^{(\sigma)} - \left[ 2 \kappa - i \left( \Delta\omega_{j}^{(b)} - \Delta\omega_{k}^{(b)} \right) \right] \mathds{1} \right\} \langle b^{\dagger}_{j} b_{k} \bm{\sigma} \rangle \nonumber \\
      &\quad\quad\quad\quad\quad\quad +
      \begin{pmatrix}
        -\frac{ 1 }{ 2 } \mathcal{E}_{j}^{*} \left( \langle b_{k} \sigma_{z} \rangle + \langle b_{k} \rangle \right) \\
        -\frac{ 1 }{ 2 } \mathcal{E}_{k} \left( \langle b_{j}^{\dagger} \sigma_{z} \rangle + \langle b_{j}^{\dagger} \rangle \right) \\
        -\gamma \langle b_{j}^{\dagger} b_{k} \rangle + \mathcal{E}_{j}^{*} \langle b_{k} \sigma_{+} \rangle + \mathcal{E}_{k} \langle b_{j}^{\dagger} \sigma_{-} \rangle
      \end{pmatrix}
      , \\
      %--------------------------------------------%
      \ddt \langle a^{\dagger}_{j} b_{k} \bm{\sigma} \rangle &= \left\{ \bm{M}^{(\sigma)} - \left[ 2 \kappa - i \left( \Delta\omega_{j}^{(a)} - \Delta\omega_{k}^{(b)} \right) \right] \mathds{1} \right\} \langle a^{\dagger}_{j} b_{k} \bm{\sigma} \rangle \nonumber \\
      &\quad\quad\quad\quad\quad\quad +
      \begin{pmatrix}
        -\frac{ 1 }{ 2 } \mathcal{E}_{j}^{*} \left( \langle b_{k} \sigma_{z} \rangle + \langle b_{k} \rangle \right) \\
        -\frac{ 1 }{ 2 } \mathcal{E}_{k} \left( \langle a_{j}^{\dagger} \sigma_{z} \rangle + \langle a_{j}^{\dagger} \rangle \right) \\
        -\gamma \langle a_{j}^{\dagger} b_{k} \rangle + \mathcal{E}_{j}^{*} \langle b_{k} \sigma_{+} \rangle + \mathcal{E}_{k} \langle a_{j}^{\dagger} \sigma_{-} \rangle
      \end{pmatrix}
      , \\
      %--------------------------------------------%
      \ddt \langle b^{\dagger}_{j} a_{k} \bm{\sigma} \rangle &= \left\{ \bm{M}^{(\sigma)} - \left[ 2 \kappa - i \left( \Delta\omega_{j}^{(b)} - \Delta\omega_{k}^{(a)} \right) \right] \mathds{1} \right\} \langle b^{\dagger}_{j} a_{k} \bm{\sigma} \rangle \nonumber \\
      &\quad\quad\quad\quad\quad\quad +
      \begin{pmatrix}
        -\frac{ 1 }{ 2 } \mathcal{E}_{j}^{*} \left( \langle a_{k} \sigma_{z} \rangle + \langle a_{k} \rangle \right) \\
        -\frac{ 1 }{ 2 } \mathcal{E}_{k} \left( \langle b_{j}^{\dagger} \sigma_{z} \rangle + \langle b_{j}^{\dagger} \rangle \right) \\
        -\gamma \langle b_{j}^{\dagger} a_{k} \rangle + \mathcal{E}_{j}^{*} \langle a_{k} \sigma_{+} \rangle + \mathcal{E}_{k} \langle b_{j}^{\dagger} \sigma_{-} \rangle
      \end{pmatrix}
      .
    \end{align}
  \end{subequations}

  %------------------------------------------------------------------------------%
  \subsection*{Third-order filter mode operators}
  %------------------------------------------------------------------------------%
  \vspace{-0.3cm}
  The operator moment equations for the third-order operators, $\langle a_{j}^{\dagger} a_{k} b_{l} \rangle, \langle b_{j}^{\dagger} b_{k} a_{l} \rangle, \langle b_{j}^{\dagger} a_{k}^{\dagger} a_{l} \rangle$, and $\langle a_{j}^{\dagger} b_{k}^{\dagger} b_{l} \rangle$ are:
  \vspace{-0.3cm}
  \begin{subequations}
    \begin{align}
      \ddt \langle a_{j}^{\dagger} a_{k} b_{l} \rangle &= -\left[ 3 \kappa - i \left( \Delta\omega_{j}^{(a)} - \Delta\omega_{k}^{(a)} - \Delta\omega_{l}^{(b)} \right) \right] \langle a_{j}^{\dagger} a_{k} b_{l} \rangle \nonumber \\
      &\quad -{ \mathcal{E}_{j}}^{*} \langle a_{k} b_{l} \sigma_{+} \rangle - \mathcal{E}_{k} \langle a_{j}^{\dagger} b_{l} \sigma_{-} \rangle - \mathcal{E}_{l} \langle a_{j}^{\dagger} a_{k} \sigma_{-} \rangle , \\
      %--------------------------------------------%
      \ddt \langle b_{j}^{\dagger} b_{k} a_{l} \rangle &= -\left[ 3 \kappa - i \left( \Delta\omega_{j}^{(b)} - \Delta\omega_{k}^{(b)} - \Delta\omega_{l}^{(a)} \right) \right] \langle b_{j}^{\dagger} b_{k} a_{l} \rangle \nonumber \\
      &\quad - \mathcal{E}_{j}^{*} \langle a_{l} b_{k} \sigma_{+} \rangle - \mathcal{E}_{k} \langle b_{j}^{\dagger} a_{l} \sigma_{-} \rangle - \mathcal{E}_{l} \langle b_{j}^{\dagger} b_{k} \sigma_{-} \rangle , \\
      %--------------------------------------------%
      \ddt \langle b_{j}^{\dagger} a_{k}^{\dagger} a_{l} \rangle &= -\left[ 3 \kappa - i \left( \Delta\omega_{j}^{(b)} + \Delta\omega_{k}^{(a)} - \Delta\omega_{l}^{(a)} \right) \right] \langle b_{j}^{\dagger} a_{k}^{\dagger} a_{l} \rangle \nonumber \\
      &\quad - \mathcal{E}_{j}^{*} \langle a_{k}^{\dagger} a_{l} \sigma_{+} \rangle - \mathcal{E}_{k}^{*} \langle b_{j}^{\dagger} a_{l} \sigma_{+} \rangle - \mathcal{E}_{l} \langle a_{k}^{\dagger} b_{j}^{\dagger} \sigma_{-} \rangle , \\
      %--------------------------------------------%
      \ddt \langle a_{j}^{\dagger} b_{k}^{\dagger} b_{l} \rangle &= -\left[ 3 \kappa - i \left( \Delta\omega_{j}^{(a)} + \Delta\omega_{k}^{(b)} - \Delta\omega_{l}^{(b)} \right) \right] \langle a_{j}^{\dagger} b_{k}^{\dagger} b_{l} \rangle \nonumber \\
      &\quad  - \mathcal{E}_{j}^{*} \langle b_{k}^{\dagger} b_{l} \sigma_{+} \rangle - \mathcal{E}_{k}^{*} \langle a_{j}^{\dagger} b_{l} \sigma_{+} \rangle - \mathcal{E}_{l} \langle a_{j}^{\dagger} b_{k}^{\dagger} \sigma_{-} \rangle .
    \end{align}
  \end{subequations}

  %------------------------------------------------------------------------------%
  \subsection*{Third-order filter mode operators with spin operators}
  %------------------------------------------------------------------------------%
  \vspace{-0.3cm}
  The operator moment equations for the fourth-order operators, $\langle a_{j}^{\dagger} a_{k} b_{l} \bm{\sigma} \rangle, \langle b_{j}^{\dagger} b_{k} a_{l} \bm{\sigma} \rangle, \langle b_{j}^{\dagger} a_{k}^{\dagger} a_{l} \bm{\sigma} \rangle$, and $\langle a_{j}^{\dagger} b_{k}^{\dagger} b_{l} \bm{\sigma} \rangle$ are:
  \vspace{-0.3cm}
  \begin{subequations}
    \begin{align}
      \ddt \langle a_{j}^{\dagger} a_{k} b_{l} \bm{\sigma} \rangle &= \left\{ \bm{M}^{(\sigma)} - \left[ 3 \kappa - i \left( \Delta\omega_{j}^{(a)} - \Delta\omega_{k}^{(a)} - \Delta\omega_{l}^{(b)} \right) \right] \mathds{1} \right\} \langle a_{j}^{\dagger} a_{k} b_{l} \bm{\sigma} \rangle \nonumber \\
      &\quad\quad\quad\quad\quad\quad +
      \begin{pmatrix}
        -\frac{ 1 }{ 2 } \mathcal{E}_{j}^{*} \left( \langle a_{k} b_{l} \sigma_{z} \rangle + \langle a_{k} b_{l} \rangle \right) \\
        -\frac{ 1 }{ 2 } \mathcal{E}_{k} \left( \langle a_{j}^{\dagger} b_{l} \sigma_{z} \rangle + \langle a_{j}^{\dagger} b_{l} \rangle \right) - \frac{ 1 }{ 2 } \mathcal{E}_{l} \left( \langle a_{j}^{\dagger} a_{k} \sigma_{z} \rangle + \langle a_{j}^{\dagger} a_{k} \rangle \right) \\
        -\gamma \langle a_{j}^{\dagger} a_{k} b_{l} \rangle + \mathcal{E}_{j}^{*} \langle a_{k} b_{l} \sigma_{+} \rangle + \mathcal{E}_{k} \langle a_{j}^{\dagger} b_{l} \sigma_{-} \rangle + \mathcal{E}_{l} \langle a_{j}^{\dagger} a_{k} \sigma_{-} \rangle
      \end{pmatrix}
      , \\
      %--------------------------------------------%
      \ddt \langle b_{j}^{\dagger} b_{k} a_{l} \bm{\sigma} \rangle &= \left\{ \bm{M}^{(\sigma)} - \left[ 3 \kappa - i \left( \Delta\omega_{j}^{(b)} - \Delta\omega_{k}^{(b)} - \Delta\omega_{l}^{(a)} \right) \right] \mathds{1} \right\} \langle b_{j}^{\dagger} b_{k} a_{l} \bm{\sigma} \rangle \nonumber \\
      &\quad\quad\quad\quad\quad\quad +
      \begin{pmatrix}
        -\frac{ 1 }{ 2 } \mathcal{E}_{j}^{*} \left( \langle a_{l} b_{k} \sigma_{z} \rangle + \langle a_{l} b_{k} \rangle \right) \\
        -\frac{ 1 }{ 2 } \mathcal{E}_{k} \left( \langle b_{j}^{\dagger} a_{l} \sigma_{z} \rangle + \langle b_{j}^{\dagger} a_{l} \rangle \right) - \frac{ 1 }{ 2 } \mathcal{E}_{l} \left( \langle b_{j}^{\dagger} b_{k} \sigma_{z} \rangle + \langle b_{j}^{\dagger} b_{k} \rangle \right) \\
        -\gamma \langle b_{j}^{\dagger} b_{k} a_{l} \rangle + \mathcal{E}_{j}^{*} \langle a_{l} b_{k} \sigma_{+} \rangle + \mathcal{E}_{k} \langle b_{j}^{\dagger} a_{l} \sigma_{-} \rangle + \mathcal{E}_{l} \langle b_{j}^{\dagger} b_{k} \sigma_{-} \rangle
      \end{pmatrix}
      , \\
      %--------------------------------------------%
      \ddt \langle b_{j}^{\dagger} a_{k}^{\dagger} a_{l} \bm{\sigma} \rangle &= \left\{ \bm{M}^{(\sigma)} - \left[ 3 \kappa - i \left( \Delta\omega_{j}^{(b)} + \Delta\omega_{k}^{(a)} - \Delta\omega_{l}^{(a)} \right) \right] \mathds{1} \right\} \langle b_{j}^{\dagger} a_{k}^{\dagger} a_{l} \bm{\sigma} \rangle \nonumber \\
      &\quad\quad\quad\quad\quad\quad +
      \begin{pmatrix}
        -\frac{ 1 }{ 2 } \mathcal{E}_{j}^{*} \left( \langle a_{k}^{\dagger} a_{l} \sigma_{z} \rangle + \langle a_{k}^{\dagger} a_{l} \rangle \right) - \frac{ 1 }{ 2 } \mathcal{E}_{k}^{*} \left( \langle b_{j}^{\dagger} a_{l} \sigma_{z} \rangle + \langle b_{j}^{\dagger} a_{l} \rangle \right) \\
        -\frac{ 1 }{ 2 } \mathcal{E}_{l} \left( \langle a_{k}^{\dagger} b_{j}^{\dagger} \sigma_{z} \rangle + \langle a_{k}^{\dagger} b_{j}^{\dagger} \rangle \right) \\
        -\gamma \langle b_{j}^{\dagger} a_{k}^{\dagger} a_{l} \rangle + \mathcal{E}_{j}^{*} \langle a_{k}^{\dagger} a_{l} \sigma_{+} \rangle + \mathcal{E}_{k}^{*} \langle b_{j}^{\dagger} a_{l} \sigma_{+} \rangle + \mathcal{E}_{l} \langle a_{k}^{\dagger} b_{j}^{\dagger} \sigma_{-} \rangle
      \end{pmatrix}
      , \\
      %--------------------------------------------%
      \ddt \langle a_{j}^{\dagger} b_{k}^{\dagger} b_{l} \bm{\sigma} \rangle &= \left\{ \bm{M}^{(\sigma)} - \left[ 3 \kappa - i \left( \Delta\omega_{j}^{(a)} + \Delta\omega_{k}^{(b)} - \Delta\omega_{l}^{(b)} \right) \right] \mathds{1} \right\} \langle a_{j}^{\dagger} b_{k}^{\dagger} b_{l} \bm{\sigma} \rangle \nonumber \\
      &\quad\quad\quad\quad\quad\quad +
      \begin{pmatrix}
        -\frac{ 1 }{ 2 } \mathcal{E}_{j}^{*} \left( \langle b_{k}^{\dagger} b_{l} \sigma_{z} \rangle + \langle b_{k}^{\dagger} b_{l} \rangle \right) - \frac{ 1 }{ 2 } \mathcal{E}_{k}^{*} \left( \langle a_{j}^{\dagger} b_{l} \sigma_{z} \rangle + \langle a_{j}^{\dagger} b_{l} \rangle \right) \\
        -\frac{ 1 }{ 2 } \mathcal{E}_{l} \left( \langle a_{j}^{\dagger} b_{k}^{\dagger} \sigma_{z} \rangle + \langle a_{j}^{\dagger} b_{k}^{\dagger} \rangle \right) \\
        -\gamma \langle a_{j}^{\dagger} b_{k}^{\dagger} b_{l} \rangle + \mathcal{E}_{j}^{*} \langle b_{k}^{\dagger} b_{l} \sigma_{+} \rangle + \mathcal{E}_{k}^{*} \langle a_{j}^{\dagger} b_{l} \sigma_{+} \rangle + \mathcal{E}_{l} \langle a_{j}^{\dagger} b_{k}^{\dagger} \sigma_{-} \rangle
      \end{pmatrix}
      .
    \end{align}
  \end{subequations}

  %------------------------------------------------------------------------------%
  \subsection*{Fourth-order filter mode operators}
  %------------------------------------------------------------------------------%
  \vspace{-0.3cm}
  The operator moment equation for the fourth-order filter mode operator $\langle a{j}^{\dagger} b_{k}^{\dagger} b_{l} a_{m} \rangle$ is:
  \vspace{-0.3cm}
  \begin{align}
    \ddt \langle a_{j}^{\dagger} b_{k}^{\dagger} b_{l} a_{m} \rangle &= -\left[ 4 \kappa - i \left( \Delta\omega_{j}^{(a)} + \Delta\omega_{k}^{(b)} \right) + i \left( \Delta\omega_{l}^{(b)} + \Delta\omega_{m}^{(a)} \right) \right] \langle a_{j}^{\dagger} b_{k}^{\dagger} b_{l} a_{m} \rangle \nonumber \\
    &\quad - \mathcal{E}_{j}^{*} \langle b_{k}^{\dagger} b_{l} a_{m} \sigma_{+} \rangle - \mathcal{E}_{k}^{*} \langle a_{j}^{\dagger} a_{m} b_{l} \sigma_{+} \rangle - \mathcal{E}_{l} \langle b_{k}^{\dagger} a_{j}^{\dagger} a_{m} \sigma_{-} \rangle - \mathcal{E}_{m} \langle a_{j}^{\dagger} b_{k}^{\dagger} b_{l} \sigma_{-} \rangle .
  \end{align}

  % End wide text for all appendix equations
\end{widetext}

%==============================================================================%
%                                 BIBLIOGRAPHY                                 %
%==============================================================================%
% Produces the bibliography via BibTeX.
\bibliography{./Bibliography}

\end{document}